\documentclass[prl,twocolumn,citeautoscript,amsmath,superscriptaddress]{revtex4-1}

\usepackage{graphicx}
\usepackage{newtxtext}
\usepackage{epsfig}
\usepackage{amsmath,bbm,amssymb,bm,times}

\usepackage[colorlinks,linkcolor=blue,citecolor=blue]{hyperref}

\begin{document}

\title{Transition from topological to chaos in the nonlinear Su-Schrieffer-Heeger model}
\author{Kazuki Sone}
\email{sone@rhodia.ph.tsukuba.ac.jp}
\affiliation{Department of Physics, University of Tsukuba, Tsukuba, Ibaraki 305-8571, Japan}
\affiliation{Department of Applied Physics, The University of Tokyo, 7-3-1 Hongo, Bunkyo-ku, Tokyo 113-8656, Japan}
\author{Motohiko Ezawa}
\affiliation{Department of Applied Physics, The University of Tokyo, 7-3-1 Hongo, Bunkyo-ku, Tokyo 113-8656, Japan}
\author{Zongping Gong}
\affiliation{Department of Applied Physics, The University of Tokyo, 7-3-1 Hongo, Bunkyo-ku, Tokyo 113-8656, Japan}
\author{Taro Sawada}
\affiliation{Department of Applied Physics, The University of Tokyo, 7-3-1 Hongo, Bunkyo-ku, Tokyo 113-8656, Japan}
\author{Nobuyuki Yoshioka}
\affiliation{Department of Applied Physics, The University of Tokyo, 7-3-1 Hongo, Bunkyo-ku, Tokyo 113-8656, Japan}
\affiliation{Theoretical Quantum Physics Laboratory, RIKEN Cluster for Pioneering Research (CPR), Wako-shi, Saitama
351-0198, Japan}
\affiliation{Japan Science and Technology Agency (JST), PRESTO, 4-1-8 Honcho, Kawaguchi, Saitama 332-0012, Japan}
\author{Takahiro Sagawa} 
\affiliation{Department of Applied Physics, The University of Tokyo, 7-3-1 Hongo, Bunkyo-ku, Tokyo 113-8656, Japan}
\affiliation{Quantum-Phase Electronics Center (QPEC), The University of Tokyo, 7-3-1 Hongo, Bunkyo-ku, Tokyo 113-8656, Japan}

\begin{abstract}
Recent studies on topological materials are expanding into the nonlinear regime, while the central principle, namely the bulk--edge correspondence, is yet to be elucidated in the strongly nonlinear regime. Here, we reveal that nonlinear topological edge modes can exhibit the transition to spatial chaos by increasing nonlinearity, which can be a universal mechanism of the breakdown of the bulk--edge correspondence. Specifically, we unveil the underlying dynamical system describing the spatial distribution of zero modes and show the emergence of chaos. We also propose the correspondence between the absolute value of the topological invariant and the dimension of the stable manifold under sufficiently weak nonlinearity. Our results provide a general guiding principle to investigate the nonlinear bulk--edge correspondence that can potentially be extended to arbitrary dimensions.
\end{abstract}
\maketitle
{\it Introduction.---}
Band topology is responsible for the existence and absence of zero modes localized at the edge of the sample, which is known as the bulk--edge correspondence \cite{Hasan2010,Qi2011}. A typical model realizing the nontrivial band topology is the Su--Schrieffer--Heeger (SSH) model \cite{Su1979}, which is a one--dimensional tight--binding model with staggered linear couplings. While band topology has been studied mainly in electronic systems such as the celebrated quantum Hall effect \cite{Klitzing1980,Thouless1982,Haldane1988}, recent studies have also revealed the existence of topological edge modes in various fields of physics including photonics \cite{Haldane2008,Khanikaev2013}, fluid \cite{Yang2015}, and cold atoms \cite{Atala2013}. Unlike the standard Schr{\"o}dinger equation, the dynamics of such classical or quantum bosonic systems are often described by nonlinear equations \cite{Gross1961,Pitaevskii1961,Marchetti2013,Luo1991,Boyd2003}. There are also attempts to extend the notion of topology to such nonlinear systems \cite{Lumer2013,Leykam2016,Harari2018,Smirnova2019,Zangeneh2019,Zhang2020,Smirnova2020,Ota2020,Ivanov2020,Lo2021,Mukherjee2021,Kotwal2021,Sone2021,Mochizuki2021,Li2022,Jurgensen2021,Mostaan2022,Wachtler2023,Isobe2023}, which have revealed that nonlinear effects can induce topological phase transitions. Especially, there are several studies on nonlinear SSH
models \cite{Chen2014,Hadad2016,Wang2019,Xia2020,Tuloup2020,Ezawa2021,Zhou2022,Engelhardt2017,Chaunsali2021,Ma2021,Jezequel2022,Ma2023}. The emergence of nonlinearity--induced topological edge modes depends on the amplitude \cite{Hadad2016,Darabi2019,Lukas2020}, based on which one can appropriately define the nonlinear topological invariants \cite{Tuloup2020,Zhou2022,Sone2023}. However, in strongly nonlinear regimes, some studies \cite{Ezawa2021,Fu2022} have pointed out the disappearance of edge modes. Thus, the bulk--edge correspondence in arbitrary strength of nonlinearity remains unelucidated.

In this paper, we reveal that the strong nonlinear effect induces the transition from topological edge modes to spatially chaotic zero modes. Such a chaos transition in zero modes is expected to be a universal mechanism of the breakdown of the bulk--edge correspondence in nonlinear systems.  Specifically, we find that the spatial distribution of zero modes is captured by a discrete dynamical system. Focusing on a minimal model of one--dimensional nonlinear topological insulators, we analyze the bifurcation in the corresponding dynamical system and reveal that it shows the period--doubling bifurcation to chaos. The bifurcation point of the period--doubling bifurcation corresponds to the parameter where the bulk--edge correspondence collapses. Concerning the physical meaning of the absolute value of a nonlinear topological invariant, we propose that under sufficiently weak nonlinearity, it corresponds to the dimension of the stable manifold in the dynamical system describing the zero modes. We demonstrate such correspondence in a model with long--range hoppings than the minimal model. Just like the minimal model, the bulk--edge correspondence of higher nonlinear topological invariants can also be broken by the chaos transition. These results can be potentially extended to arbitrary dimensions and thus provide the guiding principle to elucidate the bulk--edge correspondence and its breakdown in nonlinear topological insulators.

\begin{figure}
\includegraphics[width=86mm,bb=0 0 770 165,clip]{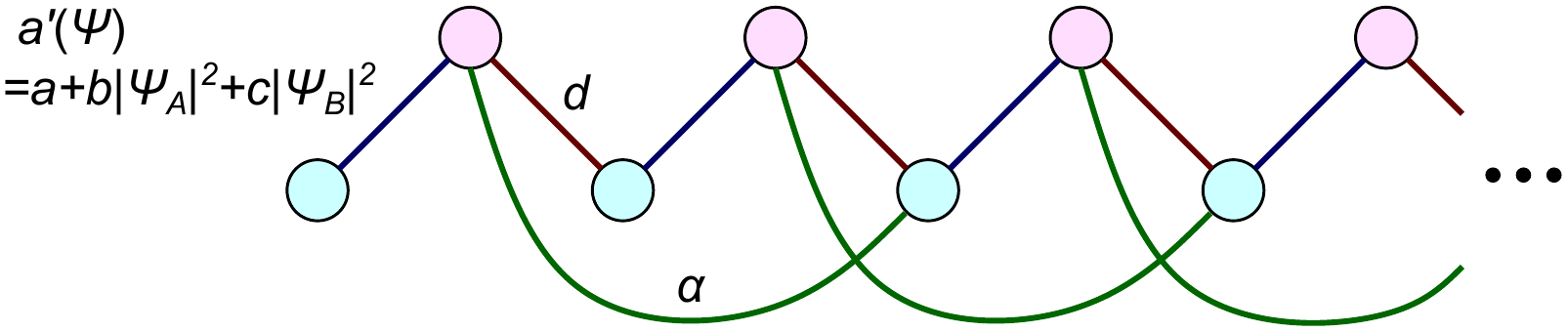}
\caption{\label{fig1}
{\bf Schematics of the long-range nonlinear SSH model.} The model has two sublattices (red and blue circles) and staggered nearest hopping. The strength of the intercell nearest--neighbor hopping depends on the state variables as $a+b|\Psi_A(x)|^2+c|\Psi_B(x)|^2$ with $a$, $b$, and $c$ being real parameters. $d$ is also a real parameter that determines the strength of intracell nearest--neighbor hopping. We also introduce long--range hoppings (green curves, the strength $\alpha$) to investigate the effect of higher winding numbers.}
\end{figure}

{\it Setup.---} Following some previous studies \cite{Bomantara2017,Tuloup2020,Zhou2022,Sone2023,Li2022}, we define the nonlinear eigenvalue problem to extend the topological invariants and edge modes to nonlinear lattice systems. Labeling the sites and internal degrees of freedom by $x$ and $j$, we consider the general dynamics $i\partial_t \Psi_j(x) = f_j(\boldsymbol{\Psi},x)$ with $f_j$ ($j=1,\cdots,M$) being nonlinear functions. Here $\boldsymbol{\Psi}$ is the abbreviation for $\{\Psi_j(x)\}_{x,j}$. We also assume the $U(1)$ symmetry, $f_j(e^{i\theta}\boldsymbol{\Psi})=e^{i\theta}f_j(\boldsymbol{\Psi})$ ($j=1,\cdots,M$), and the translation invariance, which also exist in prototypical setups of linear topological insulators \cite{Thouless1982,Haldane1988}. By imposing these symmetries, we can clearly identify the wavenumber--space description of the nonlinear system as below.

Corresponding to the nonlinear dynamics, we formulate the nonlinear eigenvalue problem by defining a nonlinear eigenvector $\boldsymbol{\Psi}$ and a nonlinear eigenvalue $E$ as a vector and a scalar satisfying $f_j(\boldsymbol{\Psi},x) = E\Psi_j(x)$. We note that the nonlinear eigenvector $\boldsymbol{\Psi}$ corresponds to a periodically oscillating steady state $\Psi_j(x;t)=e^{-iEt}\Psi_j(x)$ in the original nonlinear dynamics. Furthermore, assuming the Bloch ansatz $\boldsymbol{\Psi}(x) = e^{ikx}\boldsymbol{\psi}(k)$, one can derive the wavenumber--space description of the nonlinear eigenequation 
\begin{equation}
f_j(\boldsymbol{\psi},k) = E(k)\psi_j(k). \label{NLeigeneq}
\end{equation}
Finally, we define the nonlinear topological invariants by substituting linear eigenvectors with nonlinear ones in the definitions of conventional topological invariants. We note that a previous paper \cite{Isobe2023} discussed the bulk--edge correspondence in nonlinear eigenvalue problems with respect to eigenfrequencies, which however describe linear dynamics of higher--order differential equations. In contrast, we here focus on the situation in which the dynamics itself is nonlinear.

We next explicitly define the nonlinear winding number characterizing a one--dimensional nonlinear topological insulator with the sublattice symmetry \cite{Altland1997,Kitaev2009,Ryu2010,Jezequel2022} (see Methods) and its bulk-edge correspondence. We assume that the nonlinear eigenequation (Eq.~\eqref{NLeigeneq}) has the form of
\begin{eqnarray}
  E\left(
  \begin{array}{c}
   \boldsymbol{\psi}_A \\
   \boldsymbol{\psi}_B
  \end{array}
  \right) = 
  \left(
  \begin{array}{cc}
   0 & q(\boldsymbol{\psi},k) \\
   q^{\dagger}(\boldsymbol{\psi},k) & 0
  \end{array}
  \right)
  \left(
  \begin{array}{c}
   \boldsymbol{\psi}_A \\
   \boldsymbol{\psi}_B
  \end{array}
  \right), \label{general_sublattice}
\end{eqnarray}
where $q(\boldsymbol{\psi},k)$ is a matrix parametrized by the wavenumber $k$ and the nonlinear eigenvector $\boldsymbol{\psi}=(\boldsymbol{\psi}_A, \boldsymbol{\psi}_B)$. In this manuscript, we further focus on the case that $q(\boldsymbol{\psi},k)$ only depends on the wavenumber and the amplitude of the state $||\boldsymbol{\psi}_A||^2+||\boldsymbol{\psi}_B||^2$, $q(\boldsymbol{\psi},k)=q(||\boldsymbol{\psi}_A||^2+||\boldsymbol{\psi}_B||^2,k)$ with $||\cdot||$ being the vector norm. As in a previous study \cite{Sone2023}, we consider special solutions where the amplitude is fixed $||\boldsymbol{\psi}_A||^2+||\boldsymbol{\psi}_B||^2=w$ independently of $k$. This assumption is natural because the amplitude is a conserved quantity that one can control by tuning the energy injected to excite the initial state. Then, the nonlinear winding number is defined as
\begin{eqnarray}
 \nu(w) = \frac{1}{2\pi i} \int_0^{2\pi} \partial_k \log \left( \det q(w,k) \right) dk. \label{winding_num}
\end{eqnarray}
The amplitude dependence of the nonlinear winding number implies the potential nonlinearity--induced topological phase transition.

Regarding the bulk--edge correspondence of this nonlinear winding number, we find that the nonzero (zero) winding number basically corresponds to the existence (absence) of the localized zero modes (i.e., nonlinear eigenvectors with zero eigenvalue localized at the edge) when we identify the amplitude $w$ in Eq.~\eqref{winding_num} to be the edge amplitude, $\sum_j |\Psi_j(1)|^2=w$. We here assume the zero eigenvalue of topological edge modes because of the sublattice symmetry of the system. Unlike linear systems, one can obtain zero modes even in trivial phases, while they are anti--localized. More specifically, in semi--infinite systems, we fix the edge amplitude $||\Psi(x=1)||^2$ of the zero mode to be $w$. Then, the zero mode should exhibit localization (resp. anti--localization) in the case of $\nu(w)\neq 0$ (resp. $\nu(w)=0$). However, such a bulk--edge correspondence can be broken by the transition to chaos as we discuss later. 

Before moving to the breakdown of the bulk--edge correspondence, we discuss the triviality of the anti--localized zero modes in more detail. First, we can define a topologically trivial phase in nonlinear systems as a phase that can be adiabatically deformed into a linear trivial phase. We can show the existence of such an adiabatic deformation of the phase with the anti--localized zero mode into a linear trivial phase from the fact that one can deform the anti--localized zero mode into a diverging mode by continuously eliminating the nonlinear terms. Since such diverging modes are not regarded as physically feasible edge modes, the divergence indicates the absence of zero modes after adiabatically deforming the nonlinear system into a linear system. Therefore, the anti--localized zero modes are regarded as topologically trivial modes.

\begin{figure}
  \includegraphics[width=80mm,bb=0 0 606 546,clip]{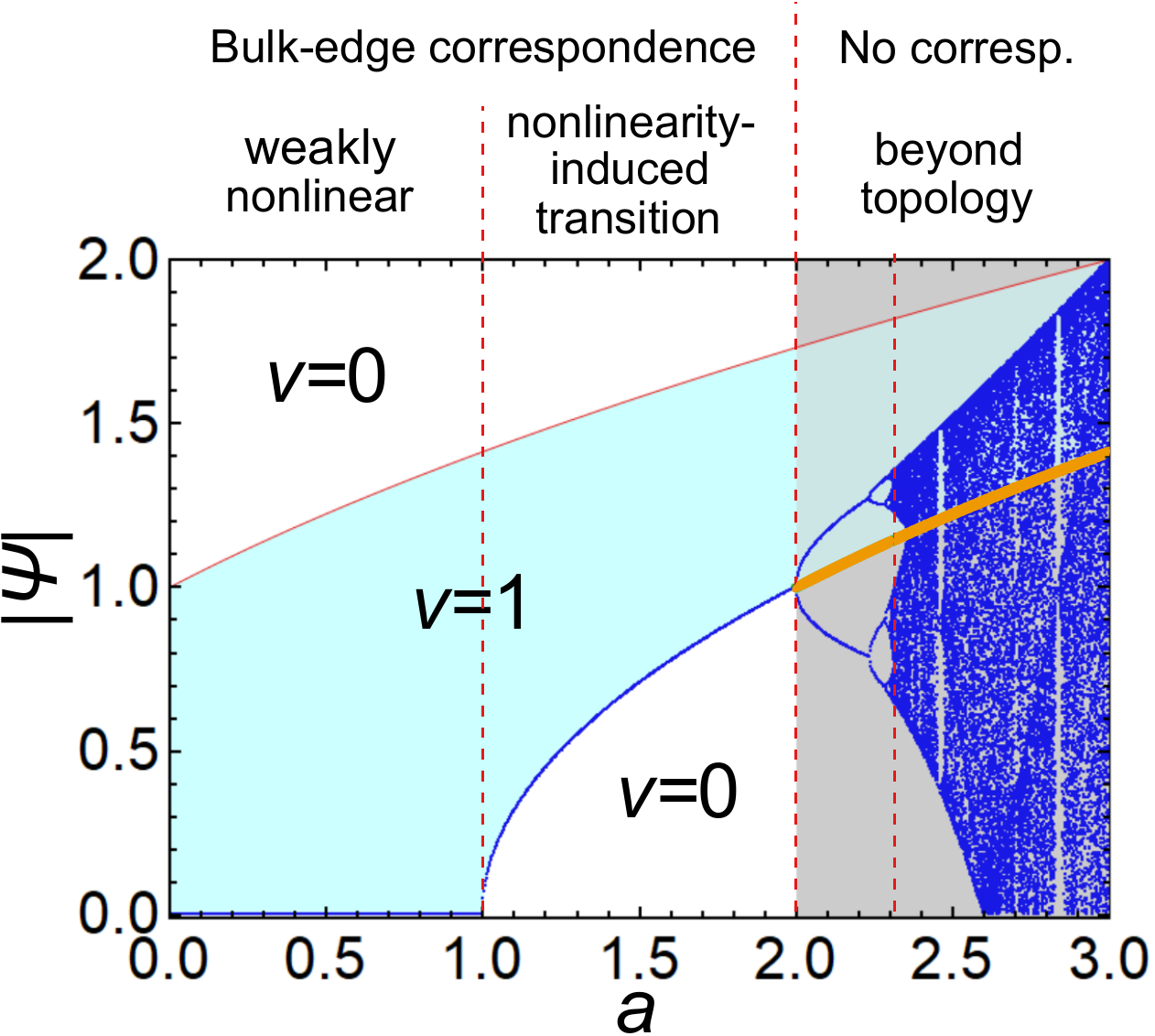}
  \caption{\label{fig2}
{\bf Bifurcation diagram of zero modes in the nonlinear SSH model.} The horizontal axis represents the parameter $a$ determining the strength of nonlinearity. The blue curves and dots represent the absolute values of zero modes $|\Psi|$ at each lattice site far from the open boundary. The red dashed lines separate the regions of weakly nonlinear topological phase, nonlinearity--induced topological phase, periodic zero mode phase, and chaotic zero mode phase. As shown in a previous paper \cite{Sone2023} for two--dimensional systems, the bulk--edge correspondence is valid from the weak to the mildly strong nonlinear regimes, by introducing the nonlinear extension of the topological number. However, in the truly strong nonlinear regime, the bulk--edge correspondence is broken down by the bifurcation and the subsequent chaos transition. The boundaries at $a=1,2$ are analytically obtained from the linear stability analysis of the spatial dynamics, and the boundary around $a=2.31$ is numerically estimated from the Lyapunov exponent. The light blue area corresponds to the parameter region where the nonlinear winding number $\nu$ becomes one, whose boundaries are analytically obtained. The upper bound of the light blue area (the red curves) corresponds to the unstable fixed points at $|\Psi_A(x)|=\sqrt{a+1}$. The orange curve shows the lower bound of the light blue area ($|\Psi_A(x)|=\sqrt{a-1}$) at $a>2$, while it does not correspond to the convergent value of the dynamical system, which indicates the breakdown of the bulk-boundary correspondence. The bulk--edge correspondence holds only at $a<2$, where steady solutions are obtained. The parameters used are $b=c=-1$, $d=1$. We set the initial condition $\Psi_A(0)=0.1$.}
\end{figure}

\begin{figure*}
  \includegraphics[width=160mm,bb=0 0 1590 585,clip]{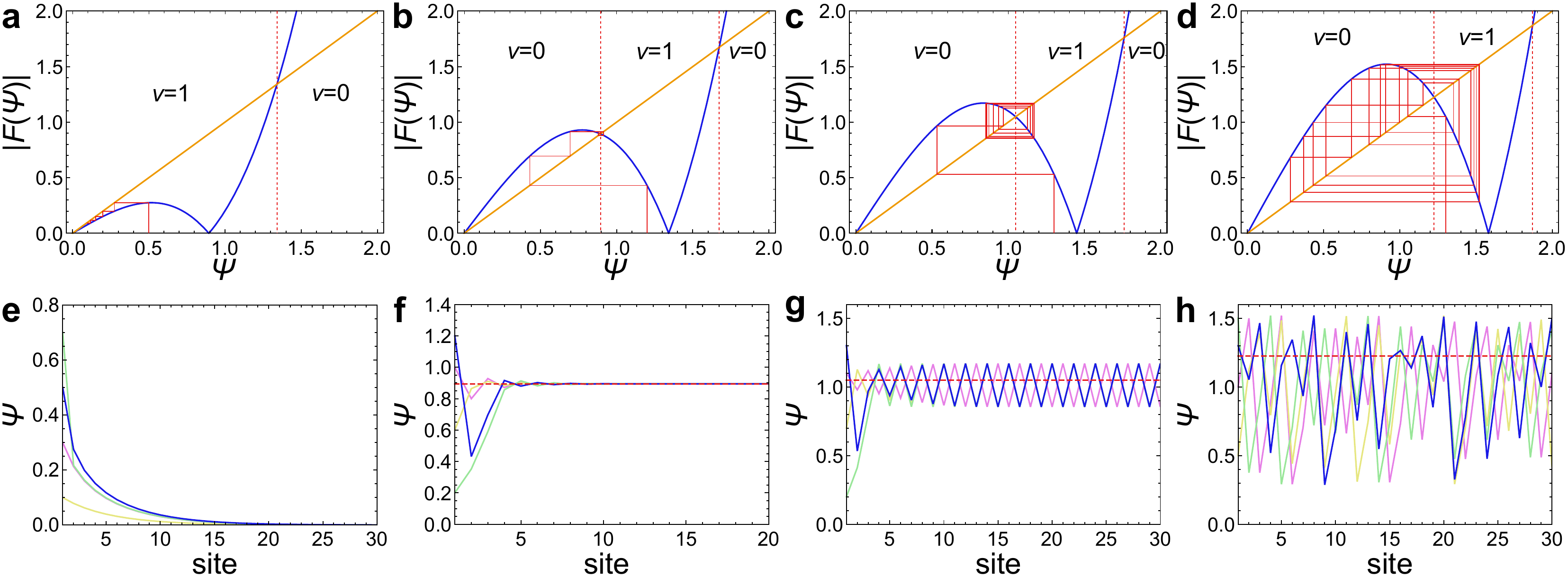}
  \caption{\label{fig3} 
  {\bf Cobweb plots and spatial distributions of zero modes in the nonlinear SSH model.} {\bf a--d} Cobweb plots at different strengths of nonlinearity $a$ are shown. The blue curves represent the absolute values of the nonlinear functions $F(\Psi)$ in Eq.~\eqref{edge_dynamics} with $\Psi$ corresponding to a component of an eigenvector. The orange lines represent $|F(\Psi)| = \Psi$. The red lines show the dynamics of Eq.~\eqref{edge_dynamics}, which correspond to the blue polylines. {\bf a} Weakly nonlinear topological case. If the nonlinear winding number is $\nu=1$ in the linear limit, localized edge modes converging to zero are obtained for smaller initial amplitudes than that represented by the red dotted line. We use the parameter $a=0.8$. {\bf b} Nonlinearity--induced topological phase. When the nonlinearity-induced topological phase transition from a trivial phase to a topological phase occurs, localized zero modes are obtained if the initial amplitude is in the region sandwiched by the red dotted lines. We use the parameter $a=1.8$. {\bf c} Periodic zero mode. We consider $a=2.1$ and obtain a stable periodic orbit. {\bf d} Chaotic zero mode. At large $a$ ($a=2.5$ in this panel), we obtain a chaotic dynamics of a zero mode. {\bf e--h} Zero modes at each parameter are shown. The colored polylines show the spatial distributions of zero modes, where the color shows the difference in the initial condition $\Psi_A(1)$. The red dashed lines show the critical amplitude of the nonlinear winding number. We use the parameters $a=0.8$, $1.8$, $2.1$, $2.5$ in panels {\bf e, f, g, h} for each, which correspond to the upper panels {\bf a--d}.} 
\end{figure*}

{\it Nonlinear SSH model.---}
To investigate the nonlinear effects on topological edge modes, we consider a nonlinear SSH model, which is a minimal model of one--dimensional nonlinear topological insulators. While previous studies have proposed various types of nonlinear extensions of the SSH model \cite{Chen2014,Hadad2016,Wang2019,Xia2020,Tuloup2020,Ezawa2021,Zhou2022,Engelhardt2017,Chaunsali2021,Ma2021,Jezequel2022,Ma2023}, we consider one of its variants that is suitable to demonstrate the chaos transition. Specifically, the nonlinear SSH model considered here has two sublattices labeled A and B (cf. Fig.~\ref{fig1}), and its dynamics is described as
\begin{eqnarray}
&{}& i\partial_t \Psi_A(x) \nonumber\\
&=& (a+b|\Psi_A(x)|^2+c|\Psi_B(x)|^2) \Psi_B(x) + d \Psi_B(x-1),\ \ \  \label{nonlinearSSH1}\\
&{}& i\partial_t \Psi_B(x) \nonumber\\
&=& (a+b|\Psi_A(x)|^2+c|\Psi_B(x)|^2) \Psi_A(x) + d \Psi_A(x+1),\ \ \  \label{nonlinearSSH2}
\end{eqnarray}
where $\Psi_{A(B)}(x)$ represents the state variables at the A (B) sublattice of the $x$th unit cell. In the following, we focus on the case that $a$, $b$, $c$, and $d$ are real, and $b$ is equal to $c$. We here consider an off--diagonal nonlinearity because it preserves the sublattice symmetry and realizes the nonlinearity--induced topological phase transition \cite{Hadad2016,Sone2023}. One can experimentally realize such an off--diagonal nonlinearity by utilizing, e.g., phase shifts by the Kerr nonlinearity in optical fibers \cite{Bisianov2019} or the nonlinear circuit elements \cite{Zhou2022} (cf. Supplementary Note 1).

To calculate the nonlinear topological invariant (Eq.~\eqref{winding_num}), we derive the wavenumber--space description of the nonlinear eigenvalue problem by assuming the Bloch ansatz. The wavenumber--space description of the nonlinear SSH model has the form of Eq.~\eqref{general_sublattice}:
\begin{eqnarray}
  E\left(
  \begin{array}{c}
   \psi_A \\
   \psi_B
  \end{array}
  \right) = 
  \left(
  \begin{array}{cc}
   0 & \tilde{a}(\psi)+de^{-ik} \\
   \tilde{a}(\psi)+de^{ik} & 0
  \end{array}
  \right)
  \left(
  \begin{array}{c}
   \psi_A\\
   \psi_B
  \end{array}
  \right), \ \ \label{eigeneq_nonlinearSSH}
\end{eqnarray}
where $\tilde{a}(\psi)$ is a function of $\psi_A$ and $\psi_B$ defined as $\tilde{a}(\psi)=a+b(|\psi_A|^2+|\psi_B|^2)$. Then, if we focus on the special solutions of nonlinear eigenvectors where $|\psi_A(k)|^2+|\psi_B(k)|^2=w$ is fixed independently of the wavenumber $k$, the nonlinear winding number becomes $\nu=\int_0^{2\pi} dk \partial_k \log (a+bw+de^{ik}) / (2\pi i)$. This nonlinear winding number becomes $\nu=1$ (resp. $\nu=0$) in the case of $a+bw<d$ (resp. $a+bw>d$), which is consistent with the linear limit $b\rightarrow0$.

\begin{figure}
  \includegraphics[width=86mm,bb=0 0 830 810,clip]{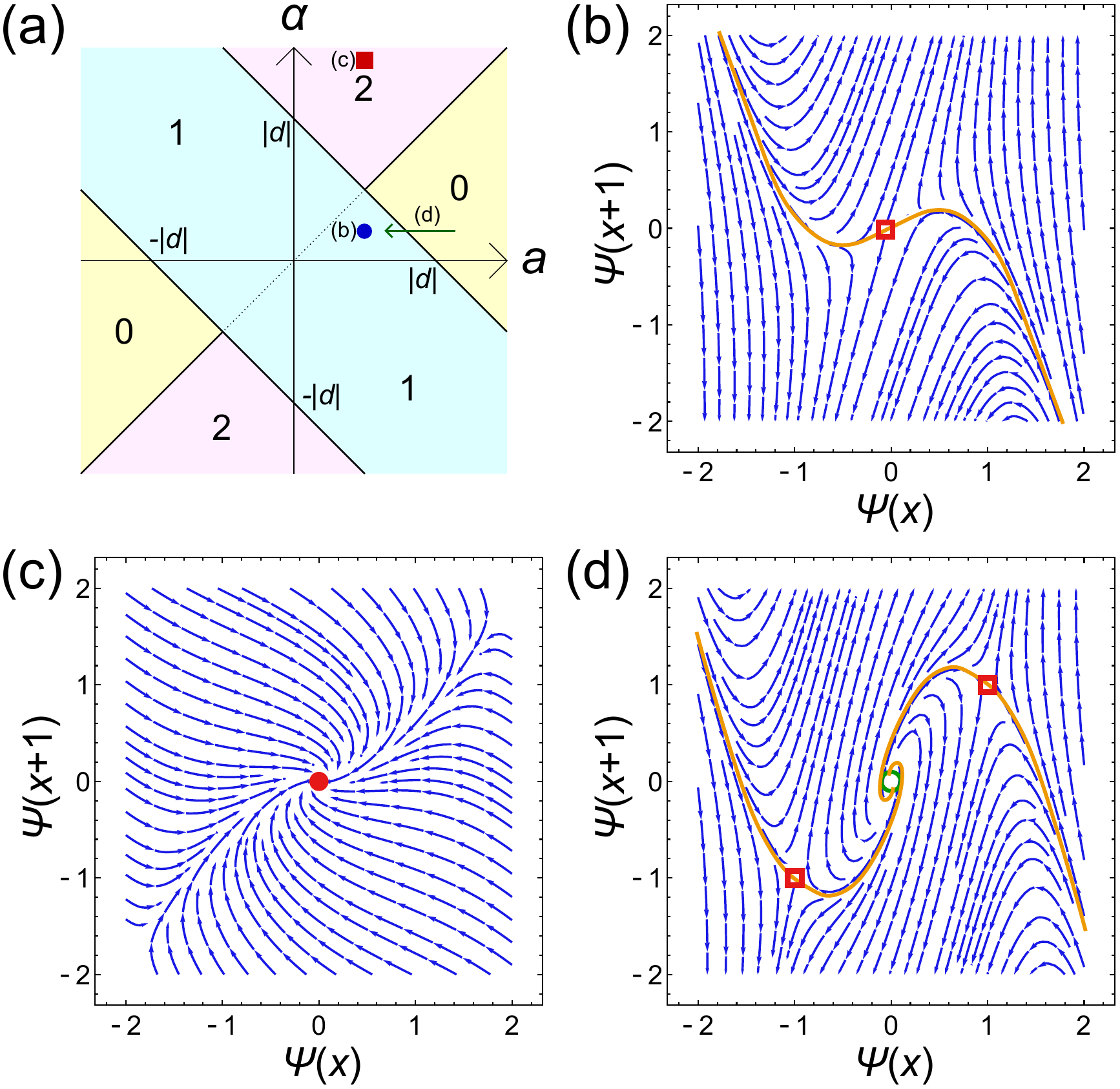}
  \caption{\label{fig4} 
  {\bf Vector fields representing the deviation of state variables in the extended nonlinear SSH model.} {\bf a} Phase diagram of the linear extended SSH model. $a$, $d$, and $\alpha$ are real parameters that determine the strength of intercell, nearest-neighbor intracell, and long-range hoppings, respectively. Each color represents the parameter regions with the same winding number, which is shown by the numbers in the regions. {\bf b--d} Vector field and the stable manifold. The blue curved arrows represent the vector field at $(\Psi(x),\Psi(x+1))$ corresponding to the values of eigenvectors at the sites $x$ and $x+1$. The red disks are stable fixed points, the red squares are saddle points, and the green circle is a fully unstable fixed point. The orange curves are eye--guides of the one--dimensional stable manifolds. {\bf b} If the winding number is one under weak nonlinearity (the blue circle in panel {\bf a}), the stable manifold is one-dimensional. The parameters used are $a=0.5$, $b=d=-1$, and $\alpha=0.25$. {\bf c} When the winding number is two under weak nonlinearity (the red filled square in panel {\bf a}), the fixed point is stable and thus has a two--dimensional stable manifold. The parameters used are $a=1$, $b=d=-1$, and $\alpha=4$. {\bf d} When the nonlinearity--induced topological phase transition from zero to one occurs (the green arrow in panel {\bf a}), nonzero fixed points appear and their stable manifold is one--dimensional. The parameters used are  $a=1.75$, $b=d=-1$, and $\alpha=0.25$.
  } 
\end{figure}

{\it Bifurcation and spatially chaotic zero modes.---} 
We analyze the zero mode of the nonlinear SSH model and reveal the breakdown of the bulk--edge correspondence between edge modes and the nonlinear winding number. Specifically, we consider the right semi-infinite system which has an open boundary at $x=1$. Then, the zero mode has zero amplitude on the B sublattice, i.e., $\Psi_B(x)=0$, and the spatial distribution on the A sublattice is described by the following nonlinear dynamical system:
\begin{equation}
\Psi_A(x+1) = -\frac{a+b|\Psi_A(x)|^2}{d} \Psi_A(x) =: F(\Psi_A(x)), \label{edge_dynamics}
\end{equation}
where $F(\Psi_A(x))$ is the nonlinear function determining the spatial distribution and independent of $\Psi_B(x)$ and $c$. As discussed above, the bulk--edge correspondence implies that the nonzero winding number corresponds to the existence of the localized zero modes, which is defined as the nonlinear eigenvector with zero eigenvalue and exhibiting a larger amplitude at the edge than in the bulk. We here calculate the zero mode from Eq.~\eqref{edge_dynamics} and confirm its localization by comparing the edge amplitude with the bulk amplitude; we judge that the zero mode is localized if $|\Psi_A(1)|>\lim_{x\rightarrow\infty}|\Psi_A(x)|$. On the other hand, the nonlinear dynamical system (Eq.~\eqref{edge_dynamics}) is known as the cubic map \cite{Rogers1983,Skjolding1983} (see Methods) and shows bifurcations to chaos. We find that this bifurcation leads to the breakdown of the bulk--edge correspondence.

We numerically demonstrate the bifurcation in the zero mode of the nonlinear SSH model by using a bifurcation diagram, which represents the behavior of the dynamical system in Eq.~\eqref{edge_dynamics} after the relaxation to a steady, periodic, or chaotic solution. In more detail, the bifurcation diagram shows the values of $\Psi_A(x)$ at large $x$ for different $a$'s (we consider $9900<x<10000$ and $0<a<3$ in Fig.~\ref{fig2}). If we obtain only one value at a fixed $a$ in the bifurcation diagram, the dynamical system (Eq.~\eqref{edge_dynamics}) converges to that value in the limit of $x\rightarrow\infty$. Before a chaos transition, the convergence of the dynamical system is destabilized at a critical value of $a$, and we obtain stable periodic solutions. Such periodic solutions are represented by two or more curves in the bifurcation diagram. After a chaos transition, one obtains scattered points at a fixed $a$, which indicates the unstable and unpredictable dynamics of chaos.

Figure \ref{fig2} shows the bifurcation diagram of Eq.~\eqref{edge_dynamics} at $b=-1$, $d=1$ (see Methods for the numerical method). At $a<2$, we confirm the convergence to a steady value. In particular, in the case of $1<a<2$, the convergent value corresponds to the critical amplitude where the nonlinear winding number is changed. This correspondence is not a coincidence, because the boundary between topological and trivial phases must be a threshold where the localization or anti--localization of a zero mode is switched, and thus a spatially uniform mode must appear at the boundary, which is a steady state of Eq.~\eqref{edge_dynamics}. In more detail, when we fix the absolute value at the boundary as $|\Psi_A(1)|^2=w$, we obtain the following equation from Eq.~\eqref{edge_dynamics}:
\begin{equation}
\Psi_A(2) = -\frac{a+bw}{d} \Psi_A(1). \label{edge_dynamics_at_boundary}
\end{equation}
Then, $|\Psi_A(2)|$ becomes larger (resp. smaller) than $|\Psi_A(1)|$ when the nonlinear winding number is $\nu=1$ (resp. $\nu=0$), which indicates the localization (resp. anti-localization) of the zero mode in a local sense. In the linear case ($b=0$), there is no $\Psi_A$ dependence in Eq.~\eqref{edge_dynamics}, and thus the winding number corresponds to the convergence to zero or the divergence. In nonlinear cases, the $\Psi_A$ dependence in Eq.~\eqref{edge_dynamics} leads to the change of rates of amplification or attenuation of a zero mode, and thus both topological and trivial zero modes can converge to a fixed value. However, such converging solutions from larger (smaller) $|\Psi_A(1)|$ than the convergent value $\lim_{x\rightarrow\infty}|\Psi_A(x)|$ still represent spatially localized (anti--localized) nonlinear eigenvectors. Since one can confirm that the convergent value $\lim_{x\rightarrow\infty}|\Psi_A(x)|$ must be larger (resp. smaller) than $|\Psi_A(1)|$ at $\nu=1$ (resp. $\nu=0$), the nonlinear winding number predicts the localization or anti-localization of the zero mode in this parameter region, which is the bulk--edge correspondence of the nonlinear winding number. As clearly seen from Eq.~\eqref{edge_dynamics}, if we set the edge amplitude $|\Psi_A(1)|^2$ to be the critical amplitude of the nonlinear winding number, $|\Psi_A(1)|^2 = w_c = -(a+d)/b$, $w_c$ also corresponds to the convergent value $w_c = \lim_{x\rightarrow\infty}|\Psi_A(x)|^2$ and we obtain a steady state where $|\Psi_A(x)|$ is independent of $x$. This is because the critical amplitude of the nonlinear winding number corresponds to the amplitude where the localization or anti--localization of a zero mode is switched.

Meanwhile, at $a>2$, the nonlinear effect destabilizes the convergence of a zero mode and induces the bifurcation to periodic solutions. Such destabilization of zero modes also breaks their localization, which induces the breakdown of the bulk--boundary correspondence. This period--doubling bifurcation ubiquitously appears before a chaos transition. In fact, we find chaotic zero modes at $a\gtrsim2.31$ by numerically confirming the positive Lyapunov exponent (see Supplementary Note 2). Since the convergence values $\lim_{x\rightarrow\infty}|\Psi_A(x)|$ are different from the bulk gap-closing points (the orange curve in Fig.~\ref{fig2}), one cannot predict such periodic and chaotic zero modes from the bulk topology, which implies the breakdown of the bulk--edge correspondence. We note that while some previous studies \cite{Engelhardt2017,Chaunsali2021,Ma2021} have discussed the instability and bifurcation in temporal dynamics of nonlinear edge modes, such temporal instability is not regarded as the breakdown of the bulk--boundary correspondence from the viewpoint of nonlinear eigenvalue problems. In contrast, the spatial chaos observed in the present model breaks the localized properties of edge modes, which induces the breakdown of the bulk-boundary correspondence unique to nonlinear systems. Interestingly, however, the temporal instability occurs at the bifurcation point in Fig.~\ref{fig2} and thus may also be related to the spatial instability (see Supplementary Note 3). We also note that since the spatial chaos is only seen in edge modes, such chaos modes are not directly relevant to the statistics of the bulk spectrum.

We also analyze the behavior of zero modes by using cobweb plots, which visualize the nonlinear discrete dynamics in Eq.~\eqref{edge_dynamics}. Figures \ref{fig3}a,e show the edge modes in the parameter region where the winding number becomes nonzero in the linear limit $b\rightarrow 0$. In such a case, we obtain a localized zero mode fully decaying to $\lim_{x\rightarrow\infty}\Psi_A(x)=0$, which corresponds to a conventional topological edge mode. If we consider the case of $1<a<2$, the amplitude of the zero mode converges to a nonzero value as shown in Figs.~\ref{fig3}b,f. Such a remaining amplitude at $x\rightarrow\infty$ is also reported in previous studies \cite{Hadad2016,Zhou2022,Sone2023} when the $w$-dependent nonlinear topological invariant signals the nonlinearity--induced topological phase transition. That is, the nonlinear winding number becomes zero (nonzero) at a smaller (larger) amplitude $w$ than the convergent value. While either amplitude gives a converging zero mode, the winding number $\nu(w)$ predicts its localization or anti--localization, which is the nonlinear bulk--edge correspondence. Considering larger $a$, we obtain periodic solutions shown in Figs.~\ref{fig3}c,g. In particular, at $x\rightarrow \infty$, the zero mode described by the red polyline takes both larger and smaller values than the initial value, and thus one cannot judge the localization or anti-localization of the zero mode. We also confirm the chaotic spatial distribution of the zero mode at $a=2.5$ (Figs.~\ref{fig3}d,h). The spatial chaos of the zero mode is also characterized by the positive Lyapunov exponent $\lambda=0.728\ldots>0$ (see also Supplementary Note 2). While we have focused on the semi--infinite system, the bulk--edge correspondence and its breakdown are also confirmed in the finite system of the nonlinear SSH model (see Supplementary Note 4). It is also noteworthy that both the edge modes and the chaos transition are robust against disorders (see Supplementary Note 5).

We emphasize that the chaos transition of zero modes can be universally seen in a wide range of nonlinear models. At the same time, such a transition is absent in some special models. For example, the SSH model with nonlinearity in the intercell hopping \cite{Hadad2016,Tuloup2020,Zhou2022} and the continuum model \cite{Sone2023} exhibit no chaos transitions, for which the bulk--boundary correspondence has been confirmed under fairly strong nonlinearity (we show such wide--range stability of edge modes in a nonlinear topological mechanics \cite{Ma2023} in Supplementary Note 6). Thus, the breakdown of the bulk--edge correspondence by chaos transitions has revealed a nontrivial nonlinear effect on topological edge modes.

{\it Extension to long-range hoppings.---}
In the previous sections, we focus on the model only with nearest--neighbor hoppings and the winding number $\nu(w)=1$ or $0$. Long--range hoppings can lead to nonlinear winding numbers larger than one. In general linear systems, the absolute value of the winding number corresponds to the number of linearly independent edge modes \cite{Kane2014}. Meanwhile, since nonlinear systems have no superposition law, it has been unclear what corresponds to the absolute value of the nonlinear winding number. We here propose that it basically corresponds to the dimension of the stable manifold in the dynamical system describing a zero mode such as Eq.~\eqref{edge_dynamics}. While previous research \cite{Ma2021} has also indicated that topological edge modes can be regarded as orbits on stable manifolds, we reveal its correspondence to nonlinear topological invariants.

To analyze the correspondence between the nonlinear winding number and the dimension of the stable manifold, we consider the extended nonlinear SSH model in Fig.~\ref{fig1}. In this model, we add next--next--to--nearest--neighbor hoppings ($\alpha \Psi_B(x-2)$ to Eq.~\eqref{nonlinearSSH1} and $\alpha \Psi_A(x+2)$ to Eq.~\eqref{nonlinearSSH2}) to the nonlinear SSH model. Figure \ref{fig4}a presents the phase diagram of the extended SSH model in the linear limit $b=c=0$ \cite{Zhang2017,Maffei2018,Hsu2020}. One can confirm that the winding number becomes $\nu=2$ at $\alpha>d-a$ and $\alpha>a$.

The dynamical system describing the spatial distribution of zero modes in the extended nonlinear SSH model reads,
\begin{equation}
\Psi_A(x+1) = -\frac{a+b|\Psi_A(x-1)|^2}{\alpha}\Psi_A(x-1) - \frac{d}{\alpha} \Psi_A(x),
\end{equation}
which looks similar to the Duffing map \cite{Guckenheimer2013} and the H{\'e}non map \cite{Henon2004}. We plot the vector field $(\Psi_A(x+1)-\Psi_A(x),\Psi_A(x+2)-\Psi_A(x+1))$ visualizing the deviations of the state variables at each step of this dynamical system in Figs.~\ref{fig4}b--d. When the winding number is one in the linear limit, the fixed point at $\Psi_A=0$ is a saddle point, and thus its stable manifold is one-dimensional. In contrast, if the winding number is two in the linear limit, the fixed point at $\Psi_A=0$ is an attractor to which any points in its neighbor converge, and thus the dimension of its stable manifold is two. These results indicate the correspondence between the nonlinear winding number and the dimension of the stable manifold in weakly nonlinear cases. We note that in weakly nonlinear cases, the dimension of the stable manifold corresponds to the number of edge modes in the linear limit, and thus one can expect the correspondence between the winding number and the dimension of the stable manifold for further higher winding numbers.

Under stronger nonlinearity, we find the breakdown of the bulk--edge correspondence in the extended nonlinear SSH model in a sense similar to the bifurcation in the original nonlinear SSH model. If we consider $-(a+\alpha)<d<0$, $0<\alpha<a$, $b<0$, and $(a+\alpha)/|d|<2$, the model exhibits a nonlinearity--induced topological phase transition and a zero mode converging to $\Psi_A(x) = (a+\alpha+d) / |b|$. In this case, the nonlinear winding number becomes $\nu=1$ and the dimension of the stable manifold is also one as shown in Fig.~\ref{fig4}d, which indicates the bulk-edge correspondence between the nonlinear winding number and the dimension of the stable manifold. However, at $(a+\alpha)/d>2$, the nonlinear winding number is unchanged, while the fixed point $\Psi_A(x) = (a+\alpha+d) / |b|$ is no longer a saddle point and becomes a fully unstable fixed point (see Supplementary Note 7). At the critical parameter $(a+\alpha)/d=2$, the bifurcation to a periodic solution occurs. Therefore, the breakdown of the bulk--edge correspondence is induced by the bifurcation as in the original nonlinear SSH model. One can also find that the long--range hopping can increase the number of anti-localized modes corresponding to the higher dimension of the stable manifold (see Supplementary Note 8).

\begin{figure}
\includegraphics[width=86mm,bb=0 0 542 260,clip]{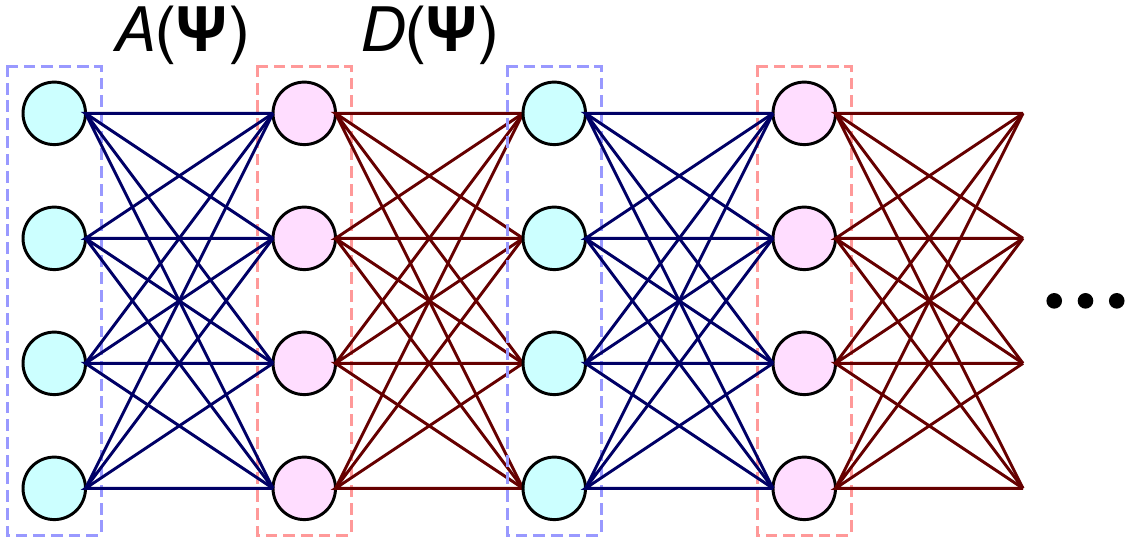}
\caption{\label{fig5} {\bf Schematic of a sublattice-symmetric one-dimensional model.} The red and blue filled circles are the sites in this model. The color of each circle represents the sublattice (the red or blue dashed square) to which the site belongs. The lines between the circles show the nonlinear hopping between them. The hopping amplitudes are alternately determined by the matrices $A$ and $D$, which depend on the nonlinear eigenvector $\boldsymbol{\Psi}$. } 
\end{figure}

{\it More general cases.---}
 While we have focused on the nonlinear SSH model, the bulk--edge correspondence can be extended to more general nonlinear systems. Specifically, we here consider a sublattice--symmetric one-dimensional model whose nonlinear eigenvalue equation is described as
\begin{widetext}
\begin{eqnarray}
  E\left(
  \begin{array}{c}
   \boldsymbol{\psi}_1 \\
   \boldsymbol{\psi}_2
  \end{array}
  \right) =
  \left(
  \begin{array}{cc}
   0 & A^{\dagger}(g(\boldsymbol{\psi}))+D^{\dagger}(g(\boldsymbol{\psi}))e^{-ik} \\
   A(g(\boldsymbol{\psi}))+D(g(\boldsymbol{\psi}))e^{ik} & 0
  \end{array}
  \right)
  \left(
  \begin{array}{c}
   \boldsymbol{\psi}_1\\
   \boldsymbol{\psi}_2
  \end{array}
  \right), \label{general_nonlinear_model}
\end{eqnarray}
\end{widetext}
where $\boldsymbol{\psi}_1$ and $\boldsymbol{\psi}_2$ are $N$--component vectors, and $A(g(\boldsymbol{\psi}))$ and $D(g(\boldsymbol{\psi}))$ are $N\times N$ matrices that are dependent on the nonlinear eigenvectors. We focus on the case that the dependence on the nonlinear eigenvector is determined only by a nonlinear function $g(\boldsymbol{\psi})$. Figure \ref{fig5} shows the schematic of this lattice model, where each site belongs to either the first or second sublattice, and the interactions only exist between different sublattices. $A(g(\boldsymbol{\psi}))$ (resp.~$D(g(\boldsymbol{\psi}))$) determines the amplitudes of intracell (resp.~intercell) couplings. Under the right semi--infinite boundary condition, a nonlinear eigenvector with zero nonlinear eigenvalue follows
\begin{equation}
 \boldsymbol{\psi}_1(x+1) = T(g(\boldsymbol{\psi}_1(x))) \boldsymbol{\psi}_1(x),\ \ \boldsymbol{\psi}_2(x)=0, \label{general_transfer}
\end{equation}
with $T$ being the state--dependent transfer matrix, which is determined by $A(g(\boldsymbol{\psi}))$ and $D(g(\boldsymbol{\psi}))$ as
\begin{equation}
T(g(\boldsymbol{\psi}_1(x))) = D^{-1}(g(\boldsymbol{\psi}_1(x))) A(g(\boldsymbol{\psi}_1(x))). \label{general_transfer_matrix}
\end{equation}
Thus, Eq.~\eqref{general_transfer} only depends on the nonlinear function $g(\boldsymbol{\psi}_1(x))$ as in the nonlinear eigenvalue problem (Eq.~\eqref{general_nonlinear_model}). Meanwhile, by assuming the Bloch ansatz $\boldsymbol{\psi}(x) = e^{ikx} \boldsymbol{\psi}(k)$ and fixing $g(\boldsymbol{\psi}(k)) = w$ independently of the wavenumber $k$ (in the original nonlinear SSH model, we set $g(\boldsymbol{\psi}(k))=||\boldsymbol{\psi}(k)||^2$), one can obtain a nonlinear winding number $\nu(w)$, which also indicates the possible usefulness of fixing nonconservative quantity. Then, if we fix the value of $g(\boldsymbol{\psi}_A(1))=w$ at the edge of the system, the dynamical system in Eq.~\eqref{general_transfer} at $x=0$ becomes $\boldsymbol{\psi}_A(2) = T(w) \boldsymbol{\psi}_A(1)$ parametrized by $w$. As is inferred from the linear cases, the nonlinear winding number $\nu(w)$ corresponds to the number of eigenvalues of $T(w)$ whose absolute values are smaller than one (such correspondence can be shown from the argument principle, Supplementary Note 9). Therefore, the correspondence between the nonlinear winding number $\nu(w)$ and amplification or attenuation around the edge site, $||\boldsymbol{\psi}(1)||<||\boldsymbol{\psi}(2)||$ or $||\boldsymbol{\psi}(1)||>||\boldsymbol{\psi}(2)||$, can be shown in more general nonlinear systems with arbitrary internal degrees of freedom. If we further assume that the tendency of the amplification or attenuation is unchanged in the limit of $x\rightarrow\infty$, one can show the bulk--edge correspondence in such a wide range of nonlinear systems. In contrast, such an assumption is broken by the chaos transition, which can also happen in this general case dependently on $T(w)$. Therefore, the breakdown of the bulk--boundary correspondence by the chaos transition also occurs in a wide range of nonlinear systems. We remain the mathematically rigorous argument as a future work.

While we have focused on sublattice-symmetric systems, topological phases and edge modes can be realized under other symmetries, such as the time--reversal and spatial--inversion symmetries. Under those symmetries, one can use the nonlinear Berry phase as a topological invariant \cite{Zhou2022} instead of the nonlinear winding number (Eq.~\eqref{winding_num}). Even without the sublattice symmetry, one can still consider spatial dynamics of edge modes similar to Eq.~\eqref{edge_dynamics}. In such spatial dynamics, it is expected that one can also observe chaos transitions, and thus the breakdown of the bulk--boundary correspondence by chaos transitions universally occurs independently of the symmetries of nonlinear systems.

We also note that there are various nonlinear systems without the $U(1)$ symmetry, such as fluids \cite{Marchetti2013}, nonlinear oscillators \cite{Luo1991}, mechanical lattices \cite{Ma2023}, and electrical circuits \cite{Zhou2022}. Such broken $U(1)$ symmetry may alter the topological classification of nonlinear systems, as is the case for interacting bosonic cases. Furthermore, a previous paper \cite{Sone2023} has shown that the $U(1)$ symmetry is necessary to obtain the wavenumber--space description of the nonlinear eigenvalue problem (cf. Eq.~\eqref{general_sublattice}) from the Bloch ansatz. Possible extension of the topological invariants to one-dimensional systems without $U(1)$ symmetry has been discussed in some previous studies \cite{Zhou2022,Zhou2024} by assuming modulated waves instead of the Bloch ansatz. By using a similar technique, we expect that our results of the present paper can be extended to more various systems even without the $U(1)$ symmetry.

{\it Discussion.---}
We revealed that nonlinear topological edge modes exhibit bifurcations to periodic solutions and chaos by analyzing the dynamical system describing the spatial distribution of zero modes. Such chaos transitions serve as the origin of the breakdown of the bulk--edge correspondence in nonlinear topology. We also proposed that the absolute value of a nonlinear topological invariant corresponds to the dimension of the stable manifold in a dynamical system describing the spatial distribution of zero modes, while such bulk--edge correspondence is also broken by the bifurcation.

While we focused on one--dimensional systems, the analytical techniques used in this paper are applicable to arbitrary dimensions. Specifically, many higher--dimensional topological insulators are reduced to low--dimensional counterparts if we fix the wavenumber to a proper value; for example, the nonlinear Qi--Wu--Zhang model studied in previous research \cite{Sone2023} is equivalent to the nonlinear SSH model if we consider the wavenumber in the $y$ direction $k_y=0,\pi$ (see Methods). By using such reduction, one can extend the arguments of chaos in nonlinear topological edge modes to higher--dimensional nonlinear topological insulators. Furthermore, even at $k_y\neq0,\pi$, one can obtain a nonlinear dynamical system in the spatial direction similar to Eq.~\eqref{edge_dynamics} (see also Methods). Therefore, the transition to chaos and the breakdown of the bulk--edge correspondence are ubiquitous in nonlinear systems. Meanwhile, our result shows that chaos, a well--known concept in nonlinear dynamical systems, can affect topological physics. Thus, we expect that interplays between nonlinear and topological physics can further uncover the characteristic behaviors in nonlinear topological insulators.

It is noteworthy that $b\neq c$ leads to the difference in the critical amplitudes at which left and right edge modes appear (see Supplementary Notes 10 and 11). Such difference can correspond to a possible $\mathbb{Z}\times\mathbb{Z}$ classification, which is reminiscent of the topological classification of one-dimensional non-Hermitian systems \cite{Gong2018,Zhou2019,Kawabata2019}. However, since we have considered conserving nonlinear dynamics that correspond to Hermitian systems, the possible $\mathbb{Z}\times\mathbb{Z}$ classification is a genuinely nonlinear effect.

There remains a possibility that one can define other topological invariants beyond the conventional wavenumber--space description and they can recover the bulk--edge correspondence in the chaotic region. Furthermore, while the models analyzed here are conservative dynamics where the sum of the amplitudes are unchanged in the time evolution, there are various dissipative (i.e., non--Hermitian--like) nonlinear systems in nature, such as biological fluids \cite{Marchetti2013} and oscillators \cite{Luo1991}. Therefore, the interplay between nonlinear and non--Hermitian topology \cite{Yuce2021,Zhu2022,Ezawa2022} and the extension of the bulk-edge correspondence to dissipative systems remain intriguing future issues.

We thank Hosho Katsura, Eiji Saitoh, and Haruki Watanabe for valuable discussions. K.S. and T. Sawada are supported by World-leading Innovative Graduate Study Program for Materials Research, Information, and Technology (MERIT-WINGS) of the University of Tokyo. K.S. is also supported by JSPS KAKENHI Grant Number JP21J20199. M.E. is supported by JST, CREST Grants Number JPMJCR20T2 and Grants-in-Aid for Scientific Research from MEXT KAKENHI (Grant No. 23H00171). Z.G. is supported by The University of Tokyo Excellent Young Researcher Program. N.Y. is supported by the Japan Science and Technology Agency (JST) PRESTO under Grant No. JPMJPR2119 and JST Grant No. JPMJPF2221. T. Sagawa is supported by JSPS KAKENHI Grant Numbers JP19H05796, JST, CREST Grant Number JPMJCR20C1, and the JST ERATO Grant Number JPMJER2302. N.Y. and T. Sagawa are also supported by Institute of AI and Beyond of the University of Tokyo.

\widetext
\pagebreak
\begin{center}
\textbf{\large Supplementary Materials}
\end{center}

\renewcommand{\theequation}{S\arabic{equation}}
\renewcommand{\thefigure}{S\arabic{figure}}
\setcounter{equation}{0}
\setcounter{figure}{0}

\subsection{Sublattice and mirror symmetries of the nonlinear Su-Scrieffer-Heeger (SSH) model.}
We here propose the definition of the symmetries in nonlinear systems and show that the nonlinear SSH model has sublattice and mirror symmetries. We start from a general nonlinear eigenequation $f_j(\boldsymbol{\Psi},x) = E\Psi_j(x)$ (Eq.~(1) in the main text). From the analogy to linear systems, we assume that the nonlinear eigenvalues under the sublattice symmetry should accompany their opposite counterparts with the same absolute values and opposite signs. If we consider a map $S$ to such an opposite counterpart, the nonlinear eigenequation should satisfy
\begin{equation}
f_j(S(\boldsymbol{\Psi}),x) = -ES(\Psi_j(x)). \label{S_eigeq}
\end{equation}
Then, if the map $S$ is linear to the constant multiple $S(a\boldsymbol{\Psi})=aS(\boldsymbol{\Psi})$, Equation \eqref{S_eigeq} reads 
\begin{equation}
S^{-1}\circ f_j\circ S(\boldsymbol{\Psi},x) = -E\Psi_j(x), \label{S_eigeq2}
\end{equation}
where $\circ$ represents the composition of a nonlinear function. Equation \eqref{S_eigeq2} is always satisfied when $f_j$ has the symmetry 
\begin{equation}
S^{-1}\circ f_j\circ S = -f_j. \label{sublattice_sym}
\end{equation}
Therefore, we adapt this equation as the definition of the sublattice symmetry. We can also define the sublattice symmetries in the wavenumber--space description as 
\begin{equation}
S^{-1}\circ f_j(k) \circ S = -f_j(k). \label{sublattice_sym_wavenumber}
\end{equation}
Equations \eqref{sublattice_sym} and \eqref{sublattice_sym_wavenumber} are derived from each other by assuming the Bloch ansatz.

We can also define the mirror symmetry in a similar way to the sublattice symmetry. Specifically, when the nonlinear eigenequation is described as $f_j(\boldsymbol{\psi}(k),k) = E(k)\psi_j(k)$, the mirror symmetry is defined as
\begin{equation}
P^{-1}\circ f_j(k) \circ P = f_j(-k), \label{mirror_sym}
\end{equation}
where $P$ is a map representing the mirror operation. The mirror symmetry guarantees that every nonlinear eigenvector appears with its parity--inversion counterpart or is inversion symmetric in itself. 

In the nonlinear SSH model, one can find the sublattice and mirror symmetries by considering linear maps $S$ and $P$. Here we denote the right--hand sides of Eq.~\eqref{eigeneq_nonlinearSSH} by ${\bf f}(k;(\psi_A,\psi_B))$. By defining $S$ as $S(\boldsymbol{\psi}(k))=\sigma_z \boldsymbol{\psi}(k)$ with $\sigma_z$ being the $z$--component of the Pauli matrix, one confirms the sublattice symmetry from the following calculation: 
\begin{eqnarray}
 &{}& S^{-1} {\bf f}(k;S(\psi_A,\psi_B)) \nonumber\\
 &=& \sigma_z {\bf f}(k;\psi_A,-\psi_B) \nonumber\\
 &=&  \left(
  \begin{array}{cc}
   1 & 0 \\
   0 & -1
  \end{array}
  \right)
  \left(
  \begin{array}{cc}
   0 & a+b|\psi_A|^2+c|(-\psi_B)|^2+de^{-ik} \\
   a+b|\psi_A|^2+c|(-\psi_B)|^2+de^{ik} & 0
  \end{array}
  \right)
  \left(
  \begin{array}{c}
   \psi_A\\
   -\psi_B
  \end{array}
  \right)  \nonumber\\
  &=& \left(
  \begin{array}{cc}
   1 & 0 \\
   0 & -1
  \end{array}
  \right) 
  \left(
  \begin{array}{c}
   -(a+b|\psi_A|^2+c|\psi_B|^2+de^{-ik}) \psi_2\\
   (a+b|\psi_A|^2+c|\psi_B|^2+de^{ik})\psi_1
  \end{array}
  \right) \nonumber\\
  &=& 
  \left(
  \begin{array}{c}
   -(a+b|\psi_A|^2+c|\psi_B|^2+de^{-ik}) \psi_2\\
   -(a+b|\psi_A|^2+c|\psi_B|^2+de^{ik}) \psi_1
  \end{array}
  \right) \nonumber\\
  &=& -{\bf f}(k;\psi_A,\psi_B).
\end{eqnarray}
Since the nonlinear SSH model holds the sublattice symmetry, one can define the nonlinear winding number. The sublattice symmetry of the nonlinear SSH model also justifies the assumption that the nonlinear eigenvalue of an edge mode is zero.

Similarly, if $b$ is equal to $c$, one can confirm the mirror symmetry of the nonlinear SSH model by defining the mirror operator as $P=\sigma_x$ ($x$--component of the Pauli matrix). In fact, we obtain
\begin{eqnarray}
 &{}& P^{-1} {\bf f}(k;P(\psi_A,\psi_B)) \nonumber\\
 &=& \sigma_x {\bf f}(k;\psi_B,\psi_A) \nonumber\\
 &=&  \left(
  \begin{array}{cc}
   0 & 1 \\
   1 & 0
  \end{array}
  \right)
  \left(
  \begin{array}{cc}
   0 & a+b|\psi_B|^2+b|\psi_A|^2+de^{-ik} \\
   a+b|\psi_B|^2+b|\psi_A|^2+de^{ik} & 0
  \end{array}
  \right)
  \left(
  \begin{array}{c}
   \psi_B\\
   \psi_A
  \end{array}
  \right)  \nonumber\\
  &=& \left(
  \begin{array}{cc}
   0 & 1 \\
   1 & 0
  \end{array}
  \right) 
  \left(
  \begin{array}{c}
   (a+b|\psi_A|^2+b|\psi_B|^2+de^{-ik}) \psi_A\\
   (a+b|\psi_A|^2+b|\psi_B|^2+de^{ik}) \psi_B
  \end{array}
  \right) \nonumber\\
  &=& 
  \left(
  \begin{array}{c}
   (a+b|\psi_A|^2+b|\psi_B|^2+de^{ik}) \psi_B\\
   (a+b|\psi_A|^2+b|\psi_B|^2+de^{-ik}) \psi_A
  \end{array}
  \right) \nonumber\\
  &=& {\bf f}(-k;\psi_A,\psi_B).
\end{eqnarray}
This mirror symmetry is related to the appearance of the gapless point at $k=0$ or $k=\pi$.

\subsection{Derivation of the dynamical system describing the spatial distribution of zero modes.} We here derive the recurrence relation in Eq.~\eqref{edge_dynamics} in the main text that describes the spatial distribution of zero modes in the nonlinear SSH model. In the right semi--infinite system, the real-space description of the nonlinear eigenequation becomes as follows:
\begin{eqnarray}
 &{}&  E\left(
  \begin{array}{c}
   \!\Psi_A(1)\! \\
   \!\Psi_B(1)\! \\
   \!\Psi_A(2)\! \\
   \!\Psi_B(2)\! \\
   \!\vdots\! \\
   \!\vdots\!
  \end{array}
  \right) = \left(
  \begin{array}{cccccc}
   0 & \!\tilde{a}(\Psi_A(1),\Psi_B(1))\! & 0 & 0 & 0 & \cdots \\
   \!\tilde{a}(\Psi_A(1),\Psi_B(1))\! & 0 & d & 0 & 0 & \cdots \\
   0 & d & 0 & \!\tilde{a}(\Psi_A(2),\Psi_B(2))\! & 0 & \cdots \\
   0 & 0 & \!\tilde{a}(\Psi_A(2),\Psi_B(2))\! & 0 & d & \cdots \\
   0 & 0 & 0 & d & 0 & \cdots \\
   \vdots & \vdots & \vdots & \vdots & \vdots & \ddots
  \end{array}
  \right)
  \left(
  \begin{array}{c}
   \!\Psi_A(1)\! \\
   \!\Psi_B(1)\! \\
   \!\Psi_A(2)\! \\
   \!\Psi_B(2)\! \\
   \!\vdots\! \\
   \!\vdots\!
  \end{array}
  \right), \nonumber \\\label{nonlinearSSH_real_semiinfinite}
\end{eqnarray}
with $\tilde{a}(\Psi_A(x),\Psi_B(x))$ being the strength of nonlinear hoppings, $\tilde{a}(\Psi_A(x),\Psi_B(x))=a+b(|\Psi_A(x)|^2+|\Psi_B(x)|^2)$. To obtain the recurrence relation (Eq.~\eqref{edge_dynamics}), we assume that the nonlinear eigenvalue is zero, $E=0$, and $\tilde{a}(\Psi_A(x),\Psi_B(x))$ is nonzero at any $x$. Then, the first row, $\tilde{a}(\Psi_A(1),\Psi_B(1))\Psi_B(1)=0$ indicates that $\Psi_B(1)=0$. By iteratively considering the equations in the odd numbers of rows, we obtain $\Psi_B(x)=0$ at any $x$. 

Next, we consider the even numbers of rows. Each of them becomes $\tilde{a}(\Psi_A(x),\Psi_B(x))\Psi_A(x) = d\Psi_A(x+1)$. Therefore, we obtain $\Psi_A(x+1) = \tilde{a}(\Psi_A(x),\Psi_B(x))\Psi_A(x)/d = [a+b(|\Psi_A(x)|^2+|\Psi_B(x)|^2]\Psi_A(x)/d$, which is equivalent to Eq.~\eqref{edge_dynamics}. Based on the real--space description of the nonlinear eigenequation (Eq.~\eqref{nonlinearSSH_real_semiinfinite}) and its reduction (Eq.~\eqref{edge_dynamics}), we calculate zero modes of the nonlinear SSH model. We note that this spatial dynamics is different from the temporal dynamics in Eqs.~\eqref{nonlinearSSH1} and \eqref{nonlinearSSH2}, and we do not calculate zero modes from the steady state of the temporal dynamics.

\subsection{Cubic map and its correspondence to the spatial dynamics in Eq.~\eqref{edge_dynamics}.}
The cubic map is a nonlinear discrete dynamics defined as
\begin{eqnarray}
 \Phi(t+1) = (1-a')\Phi(t)+a'\Phi(t)^3,
\end{eqnarray}
where $\Psi(t)$ is the state variable at time $t$. This discrete dynamics exhibits a chaos transition at $a'\sim 3.3$ \cite{Rogers1983}. One can confirm such a chaos transition from the positivity of the Lyapunov exponent. The nonlinear dynamics in Eq.~\eqref{edge_dynamics} is equivalent to the cubic map by transforming $a$ and $\Psi(x)$ as $a=(a'-1)d$ and $\Psi(x)=(-1)^x \sqrt{a'd/b}\Phi(x)$ and assuming the space index $x$ as time $t$. The chaos transition point in Fig.~\ref{fig2} is consistent with the previous study on the cubic map.

\subsection{Numerical calculation of the bifurcation diagram of the nonlinear SSH model.}
To obtain the bifurcation diagram in Fig.~\ref{fig2}, we numerically calculate the nonlinear discrete dynamics in Eq.~\eqref{edge_dynamics}. We here fix the parameters $b=-1$ and $d=1$ and change the parameter $a$. At any $a$, we set the initial condition as $\Psi_A(1) = 0.1$. To obtain $\Psi_A(x)$ at $x>1$, we iteratively calculate the right-hand side of Eq.~\eqref{edge_dynamics} and update the value of $\Psi_A(x+1)$ by using $\Psi_A(x)$. We calculate $\Psi_A(x)$ for $1\leq x \leq 10000$ at each $a$. Then, we plot the values of $\Psi_A(x)$ of $9900 < x \leq 10000$ in Fig.~\ref{fig2}. We have also used the same methods in Fig.~\ref{fig3}.

\subsection{Numerical calculation of the vector field.}
We plot the vector field $(\Psi_A(x+1)-\Psi_A(x),\Psi_A(x+2)-\Psi_A(x+1))$ in Fig.~\ref{fig4} by using the StreamPlot function in Mathematica.

\subsection{Reduction of a higher--dimensional model into a one--dimensional model.}
As is discussed in the main text, higher--dimensional models of nonlinear topological insulators are often reduced to one--dimensional models, when we consider specific wavenumbers, such as $k_y=0,\pi$. Therefore, the technique of dynamical systems used in our analysis can be extended to any dimensions, for which our result should provide a guiding principle to investigate the bulk--edge correspondence in nonlinear systems.

We here explicitly show that a two--dimensional model termed the nonlinear Qi--Wu--Zhang (QWZ) model \cite{Sone2023} is reduced to the nonlinear SSH model in this article. The nonlinear eigenvalue problem of the nonlinear QWZ model is described in the wavenumber space as
\begin{eqnarray}
 \left(
  \begin{array}{cc}
   u + \kappa w(\psi) + \cos k_x+\cos k_y  & \sin k_x + i\sin k_y \\
   \sin k_x - i\sin k_y & -[u + \kappa w(\psi) +\cos k_x+\cos k_y]
  \end{array}
  \right) \left(
  \begin{array}{c}
   \psi_A(\mathbf{k}) \\
   \psi_B(\mathbf{k})
  \end{array}
  \right) = E\left(
  \begin{array}{c}
   \psi_A(\mathbf{k}) \\
   \psi_B(\mathbf{k})
  \end{array}
  \right), \label{QWZmodel-wavenumber}
\end{eqnarray}
where $(\psi_A(\mathbf{k}),\psi_B(\mathbf{k}))^T$ is the nonlinear eigenvector at the wavenumber $\mathbf{k} = (k_x,k_y)$, and $w(\psi)=|\psi_A(\mathbf{k})|^2+|\psi_B(\mathbf{k})|^2$ represents the nonlinear term in the nonlinear QWZ model. When we focus on the wavenumber $k_y=0$, the nonlinear eigenvalue problem reads
\begin{eqnarray}
 \left(
  \begin{array}{cc}
   u + 1 + \kappa w(\psi) + \cos k_x  & \sin k_x \\
   \sin k_x & -[u + 1 + \kappa w(\psi) +\cos k_x]
  \end{array}
  \right) \left(
  \begin{array}{c}
   \psi_A(k_x,0) \\
   \psi_B(k_x,0)
  \end{array}
  \right) = E\left(
  \begin{array}{c}
   \psi_A(k_x,0) \\
   \psi_B(k_x,0)
  \end{array}
  \right). \label{QWZmodel-wavenumber2}
\end{eqnarray}
Furthermore, if we consider a unitary transformation of the nonlinear eigenvector 
\begin{equation}
\left(
  \begin{array}{c}
   \tilde{\psi}_A(k_x,0) \\
   \tilde{\psi}_B(k_x,0)
  \end{array}
  \right) = \frac{1}{\sqrt{2}}
  \left(
  \begin{array}{c}
   \psi_A(k_x,0) - i\psi_B(k_x,0) \\
   \psi_A(k_x,0) + i\psi_B(k_x,0)
  \end{array}
  \right), \label{unitary_transform}
\end{equation}
Eq.~\eqref{QWZmodel-wavenumber2} becomes
\begin{eqnarray}
 \left(
  \begin{array}{cc}
   0 & u + 1 + \kappa w(\tilde{\psi}) + e^{-ik_x} \\
   u + 1 + \kappa w(\tilde{\psi}) + e^{ik_x} & 0
  \end{array}
  \right) \left(
  \begin{array}{c}
   \tilde{\psi}_A(k_x,0) \\
   \tilde{\psi}_B(k_x,0)
  \end{array}
  \right) = E\left(
  \begin{array}{c}
   \tilde{\psi}_A(k_x,0) \\
   \tilde{\psi}_B(k_x,0)
  \end{array}
  \right), \label{QWZmodel-wavenumber3}
\end{eqnarray}
with $w(\tilde{\psi}) $ being $w(\tilde{\psi}) = |\tilde{\psi}_A(k_x,0)|^2+|\tilde{\psi}_B(k_x,0)|^2$.
Rewriting the parameters $u$, $\kappa$ as $a=u+1$, $b=\kappa$, the above eigenequation is equivalent to that of the nonlinear SSH model (Eq.~\eqref{eigeneq_nonlinearSSH}). A similar reduction is conducted at $k_y=\pi$.

For general $k_y$, the unitary transformation in Eq.~\eqref{unitary_transform} modifies the nonlinear eigenvalue problem in Eq.~\eqref{QWZmodel-wavenumber} as 
\begin{eqnarray}
 \left(
  \begin{array}{cc}
   \sin k_y & u' + \kappa w(\tilde{\psi}) + e^{-ik_x} \\
   u' + \kappa w(\tilde{\psi}) + e^{ik_x} & -\sin k_y
  \end{array}
  \right) \left(
  \begin{array}{c}
   \tilde{\psi}_A(k_x,k_y) \\
   \tilde{\psi}_B(k_x,k_y)
  \end{array}
  \right) = E\left(
  \begin{array}{c}
   \tilde{\psi}_A(k_x,k_y) \\
   \tilde{\psi}_B(k_x,k_y)
  \end{array}
  \right), \label{QWZmodel-wavenumber3}
\end{eqnarray}
with $u'$ being $u'=u+\cos k_y$. Then, by fixing $k_y$, we obtain an effectively one--dimensional nonlinear eigenequation in the real space as
\begin{eqnarray}
E \Psi_A(x) &=& \sin k_y \Psi_A(x) + (u'+\kappa|\Psi_A(x)|^2+\kappa|\Psi_B(x)|^2) \Psi_B(x) + \Psi_B(x-1), \label{nonlinear_Kane-Mele1}\\
E \Psi_B(x) &=& -\sin k_y \Psi_B(x) + (u'+\kappa|\Psi_A(x)|^2+\kappa|\Psi_B(x)|^2) \Psi_A(x) + \Psi_A(x+1), \label{nonlinear_Kane-Mele2}
\end{eqnarray}
which is a nonlinear SSH model with staggered on--site potential. The eigenvalues of nonlinear edge modes in this effectively one--dimensional system are estimated as $E=\sin k_y$ for left--localized modes and $E=-\sin k_y$ for right--localized modes.
Assuming such dispersion relation of the edge mode, we obtain a nonlinear dynamical system similar to Eq.~\eqref{edge_dynamics}. Therefore, in the nonlinear QWZ model, the chaos transition of edge modes can occur at arbitrary $k_y$. In general models, specifying the dispersion relation of edge modes can be a nontrivial problem, and thus we have to develop a way to determine the eigenvalue and eigenvector at the same time.

\subsection{Possible optical setup of the nonlinear SSH model.}
The nonlinear SSH model (Eqs. (4) and (5) in the main text) can describe various systems, such as topological photonics and electrical circuits. We here discuss the possible realization of the nonlinear coupling used in the model by using photonic systems. Specifically, we consider an interaction between two sites via an optical fiber used in Ref.~\cite{Bisianov2019}. When one uses such an optical fiber, the phase of the coupling depends on the amplitude of the light because the light propagating in the optical fiber is affected by the Kerr nonlinearity, and thus the phase shift occurs depending on its amplitude. Thus, the coupling term has a form of $c e^{i\kappa |\Psi_j|^2} \Psi_j$ with $c$ being the strength of the coupling and $\kappa$ being the coefficient of the Kerr nonlinearity. Remaining the leading order terms, we obtain $c(1+i\kappa|\Psi_j|^2-\kappa^2|\Psi_j|^4/2+\cdots)\Psi_j$. The second term has a similar form to the nonlinear coupling in the nonlinear SSH model except for the multiplication of the imaginary unit $i$. The third term is a higher--order term without the imaginary unit and thus can induce the nonlinearity--induced topological phase transition as in the models discussed in the main text.

\begin{figure}[t]
  \includegraphics[width=70mm,bb=0 0 393 282,clip]{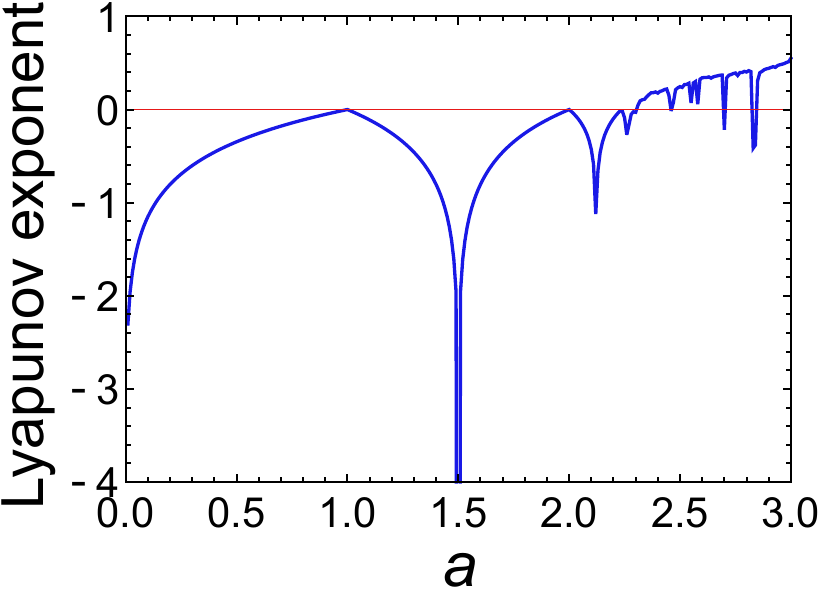}
\caption{\label{supplefig_lyapunov} {\bf Lyapunov exponents of the nonlinear SSH model.} We fix the parameters $b=c=d=1$ and change the parameter $a$. We calculate $L=10000$ steps of the dynamical system of zero modes in the nonlinear SSH model. The blue curve shows the numerically obtained Lyapunov exponents. The Lyapunov exponents above the red line are positive, which indicates the chaos of the dynamical system. 
}
\end{figure}

\subsection{Lyapunov exponents.}
Lyapunov exponents are the rate of the amplification or attenuation of the perturbation from the solution of a nonlinear dynamical system, whose positivity indicates the chaos of the dynamical system. We numerically confirm the bifurcation to chaos by calculating the Lyapunov exponent of the dynamical system describing zero modes in the nonlinear SSH model (Eq.~(7) in the main text). One can calculate the Lyapunov exponent of a one--dimensional discrete dynamical system $\psi_{i+1} = F(\psi_i)$ such as Eq.~(7) in the main text from the following formula:
\begin{equation}
 \lambda = \lim_{L\rightarrow\infty} \frac{1}{L} \sum_{i=1}^L \log |F'(\tilde{\psi}_i)|,
\end{equation}
where $F'$ is the derivative of $F$ by $\psi_i$ and $\tilde{\psi}_i$ ($i=1,\cdots,L$) is an orbit of the dynamical system $\psi_{i+1} = F(\psi_i)$.

Supplementary Figure \ref{supplefig_lyapunov} shows the numerically obtained Lyapunov exponents of the dynamical system describing zero modes in the nonlinear SSH model. We use $L=10000$ and $b=c=d=1$. Around $a=2.3$, the Lyapunov exponent becomes positive, which indicates the bifurcation to chaos. We estimate the chaos transition point in Fig.~2 in the main text from this numerical result. As is usually seen in chaotic maps, even larger $a$ than the chaos transition point, Lyapunov exponents can be negative in some parameter regions. However, in these parameter regions, the steady--state solutions are periodic and thus still break the bulk--edge correspondence.

One can also find zero Lyapunov exponents at some $a$'s. Such $a$'s correspond to the period--doubling bifurcation points. In particular, $a=1$ and $a=2$ are the bifurcation points as presented in Fig.~2 in the main text. Zero Lyapunov exponents at $a=1$ and $a=2$ can be analytically confirmed from the fact that the corresponding fixed points are linearly marginal in the spatial dynamics (Eq.~(7) in the main text), i.e., the derivative of $F$ is zero at the fixed points: $|F'(a=1,\tilde{\psi}_i=0)| = 0$ and $|F'(a=2,\tilde{\psi}_i=1)| = 0$.

\subsection{Temporal linear stability and instability of zero modes.}
While we have analyzed the stability of the spatial dynamics of zero modes in the previous section, we also conduct the linear stability analysis of the temporal dynamics of zero modes. To this end, we consider the linearization of Eqs. (4) and (5) in the main text around the periodic solution $\Psi^0_{A,B}(x) = \psi_{A,B} e^{ikx}$. We obtain the linearized dynamics,
\begin{eqnarray}
 i\frac{\partial}{\partial t} \left(
  \begin{array}{c}
   \delta\psi_A \\
   \delta\psi_B \\
   \delta\psi_A^{\ast} \\
   \delta\psi_B^{\ast}
  \end{array}
  \right) &=& 
  H_L(\psi_A,\psi_B)
 \left(
  \begin{array}{c}
   \delta\psi_A\\
   \delta\psi_B\\
   \delta\psi_A^{\ast}\\
   \delta\psi_B^{\ast}
  \end{array}
  \right)\nonumber\\
  &=& 
  \left(
  \begin{array}{cccc}
  b\psi_A^{\ast}\psi_B & \tilde{a} b|\psi_B|^2 & b\psi_A \psi_B & b\psi_B^2 \\
  \tilde{a}^{\ast} + b|\psi_A|^2 & b\psi_A\psi_B^{\ast} & b\psi_A^2 & b\psi_A \psi_B \\
  -b\psi_A^{\ast} \psi_B^{\ast} & -b\psi_B^{\ast 2} & -b\psi_A\psi_B^{\ast} & -\tilde{a}^{\ast} - b|\psi_B|^2 \\
  -b\psi_A^{\ast 2} & -b\psi_A^{\ast} \psi_B^{\ast} & -\tilde{a} - b|\psi_A|^2 & -b\psi_A^{\ast}\psi_B
  \end{array}
  \right)
 \left(
  \begin{array}{c}
   \delta\psi_A\\
   \delta\psi_B\\
   \delta\psi_A^{\ast}\\
   \delta\psi_B^{\ast}
  \end{array}
  \right),
\end{eqnarray}
where we consider a periodic modulation $\Psi_{A,B}(x) = \Psi^0_{A,B}(x)+\delta \psi_{A,B} e^{ik' x}$ and $\tilde{a} = a -b(|\psi_A|^2+|\psi_B|^2)+ de^{ik'}$. We note that the wavenumber of this periodic modulation can be different from that of the periodic solution $\Psi^0_{A,B}(x)$.

By calculating the eigenvalues of $H_L(\psi_A,\psi_B)$, we find that the bifurcation in the nonlinear dynamical system in the spatial direction (Eq. (7) in the main text) is also related to the temporal instability of zero modes. We here consider the periodic solution corresponding to bulk modes with zero eigenvalues: $k=\pi$, $(\psi_{A},\psi_B) = (w/\sqrt{2},\pm w/\sqrt{2})$, $w = a-1$. Then, the eigenvalues of $H_L(\psi_A,\psi_B)$ become $\lambda_{1\pm} = \pm \sqrt{|a+de^{ik'}|^2+2bw(a+d\cos k')+(bw)^2}$ and $\lambda_{2\pm} = \pm\sqrt{|a+de^{ik'}|^2+4bw(a+d\cos k')+3(bw)^2}$. $\lambda_{1\pm}$ are always real, while $\lambda_{2\pm}$ become purely imaginary if $|a+de^{ik'}|^2+4bw(a+d\cos k')+3(bw)^2$ is negative. Since $|a+de^{ik'}|^2+4bw(a+d\cos k')+3(bw)^2$ takes minimum at $k'=0$ or $k'=\pi$ and becomes zero at $k'=\pi$, $\lambda_{2\pm}$ are real if $(a+d)^2+4bw(a+d)+3(bw)^2$ is positive. In the case of negative $b$, we obtain the condition for the linear instability, i.e., imaginary $\lambda_{2\pm}$ as $(a+d)/(3|b|)<w<(a+d)/|b|$. When we assume the parameters $a>0$, $b=-1$, and $d=1$ as in Fig.~2 in the main text, one can confirm that this inequality is equivalent to $a>2$, at which the localized zero mode becomes unstable in the spatial dynamics in Eq.~(7) in the main text. Thus, the spatial and temporal instability of zero modes are closely related to each other.

\subsection{Bulk-edge correspondence in finite systems.} 
We here numerically confirm the bulk--edge correspondence in finite systems of the nonlinear SSH model (Eqs.~(4) and (5) in the main text). To judge the localization or anti--localization of zero modes, we define the indicator of localization $P$ as 
\begin{eqnarray}
 P &=& w_{\rm edge}/w_{\rm second},\label{localization_indicator}\\
 w_{\rm edge} &=& \max \left( \sum_j \Psi_j(1), \sum_j \Psi_j(L) \right), \\
 w_{\rm second} &=&  \begin{cases}
\sum_{j} |\Psi_j(2)|^2 & \sum_j |\Psi_j(1)|^2 > \sum_j|\Psi_j(L)|^2 \\
\sum_{j} |\Psi_j(L-1)|^2 & ({\rm otherwise})
\end{cases},
\end{eqnarray}
with $L$ being the system size. $\Psi_j(x)$ is the nonlinear eigenvector of a zero mode and $x$ and $j$ represent the location and the internal degrees of freedom, respectively. This localization indicator becomes larger (smaller) than one, when the zero mode is localized (anti--localized) at the edge. We note that one should modify the definition of the localization indicator in long--range models, while we here focus on the nonlinear SSH model whose hopping range is one.

\begin{figure*}[t]
  \includegraphics[width=160mm,bb=0 0 912 234,clip]{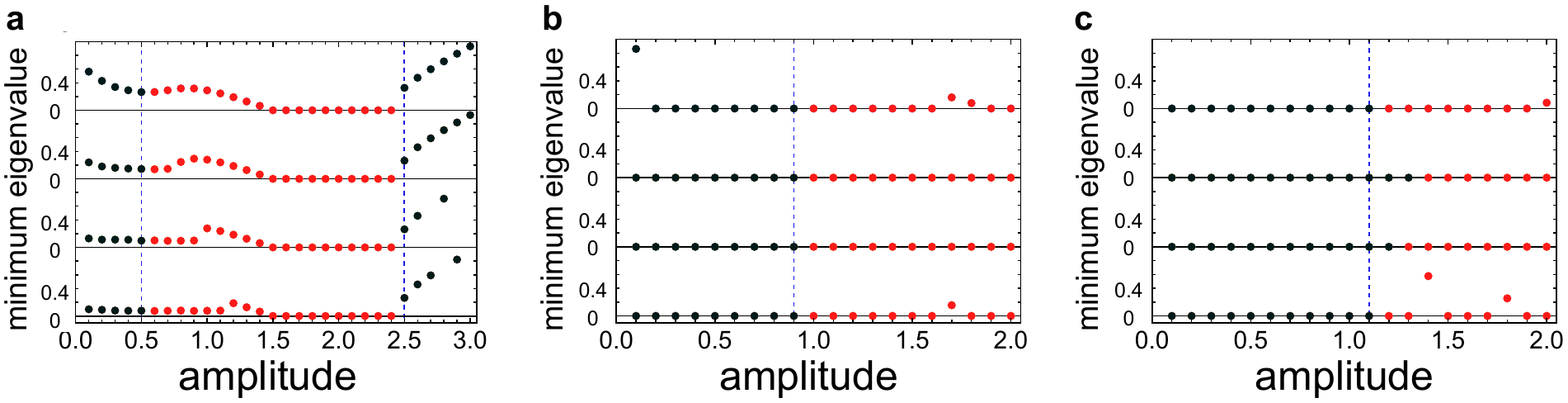}
\caption{\label{supplefig_indicator} {\bf Minimum absolute values of eigenvalues and localization indicators $P$ of the finite system of the nonlinear SSH model.} We fix the parameters $b=c=d=1$. We plot the minimum absolute values of the nonlinear eigenvalues at the system size $L=5$, $10$, $15$, and $20$ from the top. The red (black) circles represent the localization $P>1$ (anti--localization $P<1$) of the corresponding nonlinear eigenvectors. {\bf a} Localization indicators at $a=1.5$. The blue dashed lines are the transition point of the nonlinear winding number. We can confirm the bulk--edge correspondence between the nonlinear winding number and the localized zero modes. We note that some data points disappear at $a>2.5$ due to the difficulty of the convergence of the numerical techniques under the strong nonlinearity. {\bf b} Localization indicators at $a=1.9$. The localization and anti--localization of zero modes are switched at the transition point represented by the blue dashed line, which indicates the bulk--edge correspondence. {\bf c} Localization indicators at $a=2.1$. Even at the amplitude larger than the critical amplitude $w=1.1$ (the blue dashed line), there exist anti--localized zero modes. Therefore, the bulk--edge correspondence is broken in this parameter regime, which is consistent with the analysis in the semi--infinite system.}
\end{figure*}

\begin{figure*}[t]
  \includegraphics[width=120mm,bb=0 0 645 280,clip]{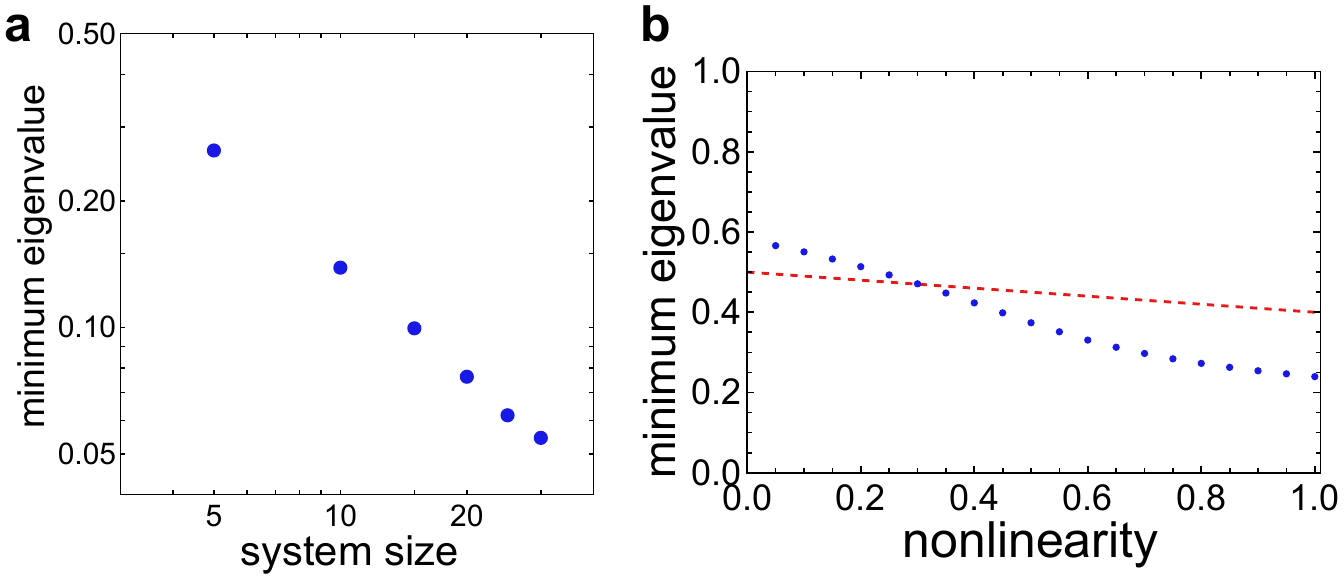}
\caption{\label{supplefig_scaling} {\bf Size-- and nonlinearity--dependence of the minimum absolute values of nonlinear eigenvalues.} {\bf a} Finite-size scaling of the minimum absolute values of nonlinear eigenvalues in the nonlinear SSH model. We use the parameters $a=0.6$ and $b=c=d=1$. We confirm that the minimum eigenvalues are inversely proportional to the system size and thus converge to zero in the thermodynamic limit. {\bf b} Disappearance of the anti--localized zero mode in the linear limit. The red dashed line shows the bulk band gap obtained from the Bloch ansatz. The horizontal axis shows the value of $b$ (we set $b=c$). We fix the other parameters as $a=1.5$, $d=1$, and the amplitude $w=0.1$. At $b=0.3$, the minimum absolute values of the nonlinear eigenvalues become larger than the bulk band gap. Thus, one cannot distinguish the anti--localized modes and bulk modes at $b<0.3$, which implies the disappearance of the anti--localized zero modes in the linear limit.}
\end{figure*}

We numerically solve the nonlinear eigenvalue problem of the finite chains of the nonlinear SSH model with different system sizes $L=5$, $10$, $15$, and $20$. To solve the nonlinear eigenvalue problem, we assume the eigenequation as an algebraic equation and use the quasi--Newton method \cite{Sone2023}. Supplementary Figure \ref{supplefig_indicator} shows the minimum absolute values of eigenvalues and the localization indicators (Supplementary Eq.~\eqref{localization_indicator}) of the corresponding eigenvectors. At $a=1.5$ (Supplementary Fig.~\ref{supplefig_indicator}a), the nonlinear winding number is nonzero $\nu_{\rm NL}=1$ in the range of amplitude, $0.5<w<2.5$. In this regime, we obtain localized zero modes corresponding to the nonzero winding number. Meanwhile, we obtain anti--localized zero modes at $w<0.5$ and no zero modes at $w>2.5$. Therefore, the nonlinear winding number predicts the existence or absence of the localized zero modes at $a=1.5$. We can also confirm the bulk--edge correspondence at $a=1.9$ as shown in Supplementary Fig.~\ref{supplefig_indicator}b.

In contrast, if we consider $a=2.1$, we obtain anti--localized zero modes even in the topologically nontrivial phase at $w>1.1$ (Supplementary Fig.~\ref{supplefig_indicator}c). Therefore, the bulk--edge correspondence is broken in this parameter regime, which is consistent with the result in the semi--infinite system shown in Fig.~2 in the main text. We note that the zero modes in the semi--infinite system exhibit the localization indicator $P>1$ even at $a>2$ and thus can be regarded as locally localized but globally anti--localized zero modes. Meanwhile, the zero modes in finite systems can be locally anti--localized in the sense that the localization indicator $P$ is smaller than one. While both the breakdowns of the bulk--edge correspondence observed in finite and semi--infinite systems are related to the period--doubling bifurcation, the detailed mechanism of the anti--localization of zero modes in finite systems remains a future issue.

The obtained nonlinear eigenvalues are not exactly zero due to the finite--size effect. To confirm the convergence of nonlinear eigenvalues to zero in the thermodynamic limit $L\rightarrow\infty$, we investigate the finite--size scaling of the minimum absolute values of nonlinear eigenvalues. Supplementary Figure \ref{supplefig_scaling}a shows the minimum absolute values of nonlinear eigenvalues at different system sizes. We here use the parameters $a=1.5$, $b=c=1$, and $d=1$, and fix the edge amplitude to be $w_{\rm edge}=0.6$. We confirm that the nonlinear eigenvalues are inversely proportional to the system size and thus converge to zero in the thermodynamic limit. This system--size dependence of the nonlinear eigenvalue is different from the linear case, which is exponentially decreased in the system size. This is because the nonlinearity--induced edge modes have nonvanishing amplitudes far from the edge $x\rightarrow\pm\infty$, and the left and right localized modes $\Psi_{\rm left}$, $\Psi_{\rm right}$ have a larger resonance integral than in linear cases. In fact, the resonance integral becomes
\begin{equation}
\frac{ \sum_{x=1}^L \sum_{j=A,B} \Psi_{{\rm left},j}(x)^{\ast} \vec{f}(\Psi_{{\rm right},j}(x))}{\sqrt{\sum_{x=1}^L \sum_{j=A,B} \Psi_{{\rm left},j}(x)^{\ast} \Psi_{{\rm left},j}(x)}\sqrt{\sum_{x=1}^L \sum_{j=A,B} \Psi_{{\rm right},j}(x)^{\ast} \Psi_{{\rm right},j}(x)}} = \frac{\mathcal{O}(1)}{\mathcal{O}(L)} = \mathcal{O}(L^{-1}),
\end{equation}
where $\vec{f}$ represents the right--hand side of Eqs.~(4) and (5) in the main text. To obtain the system--size dependence, we utilize the fact that $\vec{f}(\Psi_{{\rm right},j}(x))$ is zero in the bulk, and thus 
\begin{equation}
\sum_{x=1}^L \sum_{j=A,B} \Psi_{{\rm left},j}(x)^{\ast} \vec{f}(\Psi_{{\rm right},j}(x)) = \sum_{j=A,B} \left[ \Psi_{{\rm left},j}(1)^{\ast} \vec{f}(\Psi_{{\rm right},j}(1)) + \Psi_{{\rm left},j}(L)^{\ast} \vec{f}(\Psi_{{\rm right},j}(L))\right]
\end{equation}
becomes almost independent of the system size. Thus, the interaction between the left and right localized modes is also proportional to the inverse of the system size, which results in the power--law system--size dependence of the nonlinear eigenvalue of edge modes. We note that similar size dependence is observed in a previous paper on the nonlinear Chern number \cite{Sone2023}.

It is also noteworthy that the anti--localized zero modes disappear in the linear limit $b,c\rightarrow 0$. We consider the case of $b=c$ and numerically confirm such disappearance of the anti--localized zero modes by calculating the nonlinear eigenvalues at different $b$. Supplementary Figure \ref{supplefig_scaling}b shows the minimum absolute values of the nonlinear eigenvalues and the size of the bulk band gaps predicted from the Bloch ansatz. At sufficiently small $b$, the nonlinear eigenvalues become larger than the bulk band gaps, and thus one cannot distinguish the anti--localized zero modes and bulk modes. Therefore, anti--localized zero modes are induced by the nonlinearity and disappear in the linear limit.

\begin{figure*}[t]
  \includegraphics[width=140mm,bb=0 0 912 475,clip]{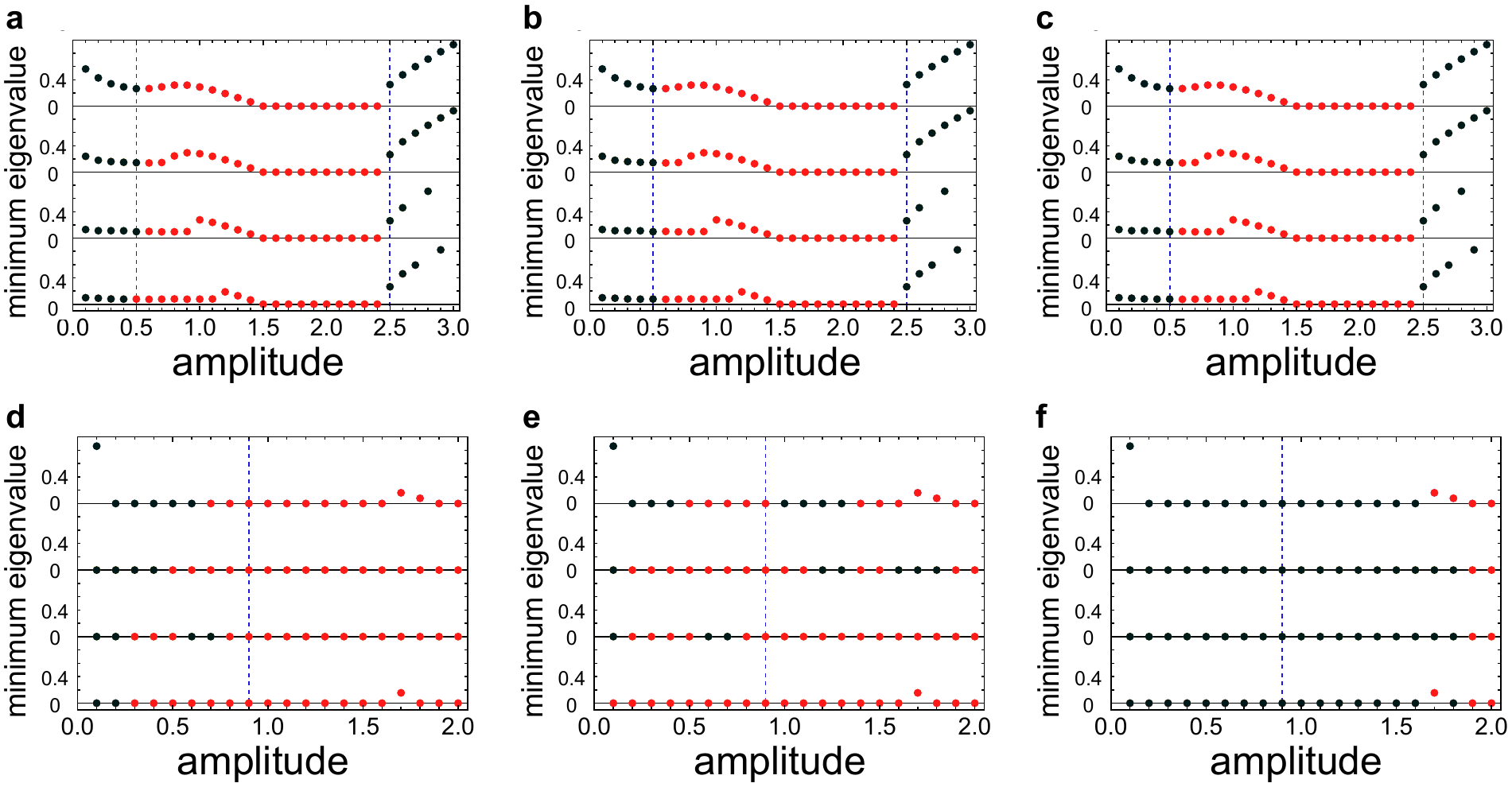}
\caption{\label{supplefig_indicator_other} {\bf  Minimum absolute values of eigenvalues and different definitions of localization indicators $P_{1,2,3}$ of the finite system of the nonlinear SSH model.} We fix the parameters $b=c=-1$ and $d=1$. We plot the minimum absolute values of the nonlinear eigenvalues at the system size $L=5$, $10$, $15$, and $20$ from the top. The red (resp. black) circles represent $P_{1,2,3}>1$ (resp. $P_{1,2,3}<1$) of the corresponding nonlinear eigenvectors. {\bf a-c} Localization indicators at $a=1.5$. Panels {\bf a}, {\bf b}, and {\bf c} show the localization indicators $P_1$, $P_2$, and $P_3$, respectively.
The blue dashed lines are the transition point of the nonlinear winding number. We can confirm the bulk--edge correspondence between the nonlinear winding number and the localized zero modes. We note that some data points disappear at an amplitude larger than $2.5$ due to the difficulty of the convergence of the numerical techniques under the strong nonlinearity. In this parameter region, all the indicators correspond to the bulk topological invariants. {\bf d-f} Localization indicators at $a=1.9$. Panels {\bf d}, {\bf e}, and {\bf f} show the localization indicators $P_1$, $P_2$, and $P_3$, respectively. The blue dashed lines show the transition point of the nonlinear winding number, while the localization indicators are unchanged at those boundaries. Since the bulk--edge correspondence should still be valid in this parameter region as inferred from the bifurcation diagram in Fig.~2 in the main text, the localization indicators $P_{1,2,3}$ are useless to discuss the bulk--boundary boundary correspondence in finite systems.}
\end{figure*}

While we have used the localization indicator $P$ (Supplementary Eq.~\eqref{localization_indicator}) in the above numerical calculations, there are other possible definitions of the localization indicators. Specifically, we can consider the following definitions that distinguish the localization and anti--localization of zero modes $\Psi_j(x)$ in a global sense:
(1) $P_1=w_{\rm edge}/w_{\rm ave}$ with $w_{\rm ave}=\sum_{x,i} |\Psi_i(x)|^2/L$ being the averaged amplitude, (2) $P_2=w_{\rm edge}/w(x)$ with $w(x)=\sum_{i} |\Psi_i(x)|^2$ being the amplitude and $x$ denoting a bulk site that we determine in hand, (3) $P_3=w_{\rm edge}/w_{\rm max}$ with $w_{\rm max}=\max_x \left( \sum_{i} |\Psi_i(x)|^2 \right)$ being the maximum amplitude.
Unfortunately, these localization indicators can be inconsistent with the nonlinear winding number at $a<2$, where the bulk--edge correspondence still holds true in the nonlinear SSH model. Supplementary Figure \ref{supplefig_indicator_other} shows the localization indicators of the nonlinear eigenvectors with the minimum absolute values of nonlinear eigenvalues. At $a=1.5$, all the indicators correspond to the localization and anti--localization of zero modes predicted from the nonlinear winding number. In contrast, at $a=1.9$, the indicators do not correspond to the nonlinear winding number. In this parameter regime, the bulk--edge correspondence is confirmed in semi--infinite systems (Fig.~2 in the main text). Therefore, the localization indicators $P_1$, $P_2$, and $P_3$ are unsuitable to discuss the bulk--edge correspondence in finite nonlinear systems.

\subsection{Robustness of edge modes and chaos transitions against spatial disorders}
We here confirm that topological edge modes and their chaos transitions are robust against disorders in the nonlinear SSH model. For simplicity, we consider the spatial inhomogeneity of $a$ and $d$ and fix the other parameters $b=c=-1$ independently of the location. We note that if we introduce disorders in $b$ and $c$, we can renormalize their effects into those of $a$ and $d$. We introduce the disorder in $a$, $d$ by setting $a(x)$, $d(x)$ at each location $x$ as $(1+\Delta a(x))a_{\rm ave}$, $(1+\Delta d(x))d_{\rm ave}$, where $\Delta a(x)$ and $\Delta d(x)$ are randomly determined from a uniform distribution $\left[-\delta,\delta \right]$. We fix the expectation value of $d(x)$ as $d_{\rm ave}=1$. We calculate the spatial dynamics similar to Eq.~(7) in the main text at $\delta=0.01$ and $0.05$.

\begin{figure*}[t]
  \includegraphics[width=120mm,bb=0 0 1305 895,clip]{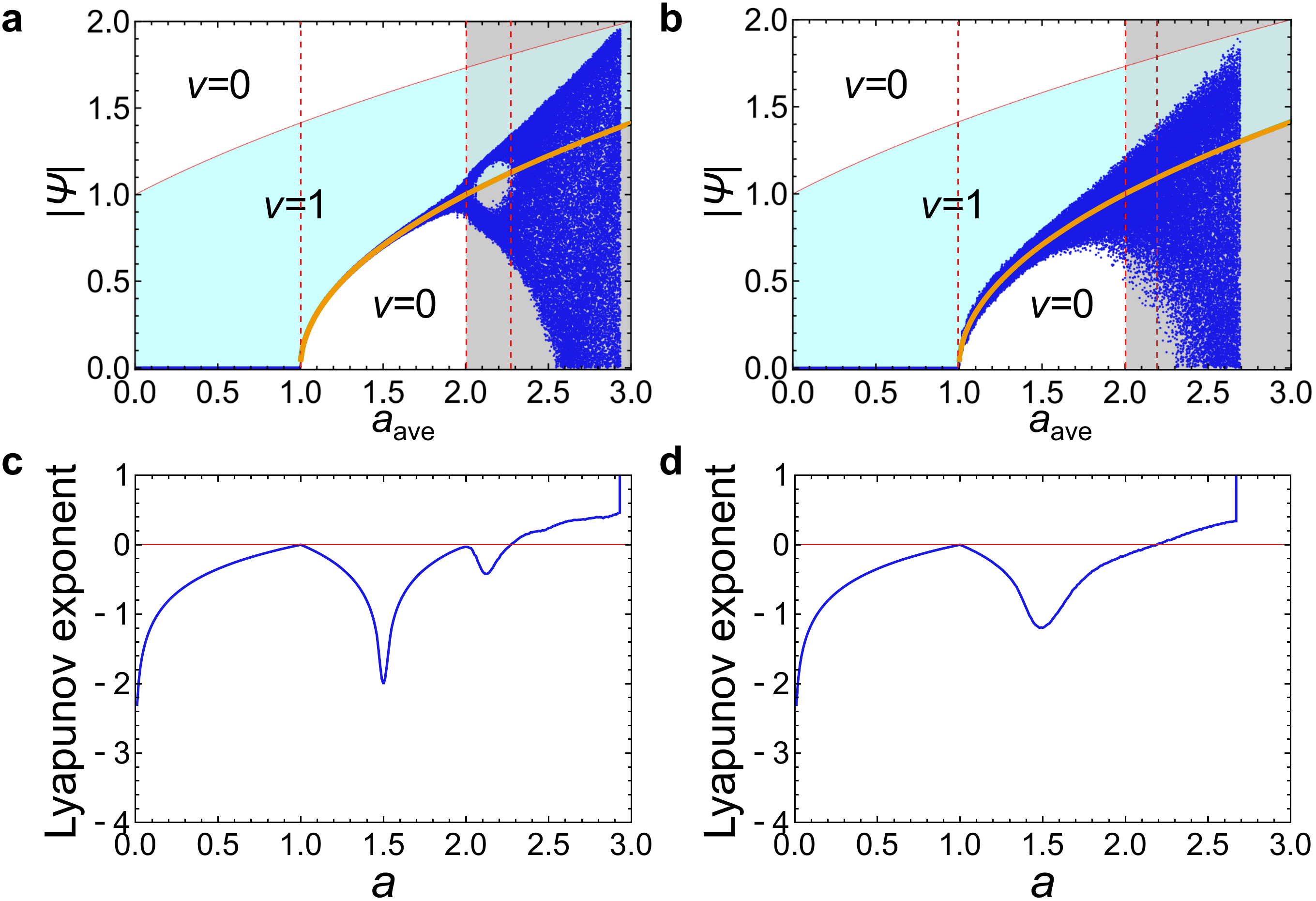}
\caption{\label{supplefig_disorder} {\bf Bifurcation plots and Lyapunov exponents under the existence of disorder.} {\bf a,b,} Bifurcation plots under the existence of disorder. We fix the parameters $b=c=-1$ and $d_{\rm ave}=1$, and change $a_{\rm ave}$. We set the strength of disorder as $\delta=0.01$ in panel {\bf a} and $\delta=0.05$ in panel {\bf b}. The red lines at $a_{\rm ave}=1$ and $a_{\rm ave}=2$ represent the bifurcation points in the disorder-free case. The third red line shows the chaos transition point obtained from the numerical calculation of the Lyapunov exponents in panels {\bf c} and {\bf d}. We obtain almost convergent solutions at $a_{\rm ave}<2$, which indicates the robustness of topological edge modes against disorders. One can confirm the period--doubling bifurcation around $a_{\rm ave}\sim2$ in panel {\bf a}, while the other period--doubling bifurcations seen in the disorder--free case disappear. We note that at large $a_{\rm ave}$, the spatial dynamics diverge to infinity, and thus there are no data points.
{\bf c,d,} Lyapunov exponents under the existence of disorder. We use the same parameters as in panels {\bf a} and {\bf b} and set the strength of disorder as $\delta=0.01$ in panel {\bf c} and $\delta=0.05$ in panel {\bf d}. For both strengths of disorders, we obtain positive Lyapunov exponents (above the red lines) at any larger $a_{\rm ave}$ than critical values, which indicates the robustness of the chaos transition against disorder. At large $a_{\rm ave}$, since the spatial dynamics diverge to infinity, the Lyapunov exponents also diverge.
}
\end{figure*}

Supplementary Figure \ref{supplefig_disorder} shows the bifurcation diagram at each strength of the disorder. At $a_{\rm ave}<2$, the convergent behavior to $|\Psi_A(x)|\sim \sqrt{(a-b)/d} $ seems to remain if we permit the fluctuation of $\Psi_A(x)$ comparable to that of $a(x)$. We find that the period--doubling bifurcations disappear at large disorder $\delta$. At large $a_{\rm ave}$, the spatial dynamics becomes very noisy, which implies the spatial chaos of zero modes.

We also calculate the Lyapunov exponents under the existence of disorder (Supplementary Fig.~\ref{supplefig_disorder}c,d) as in Supplementary Fig.~\ref{supplefig_lyapunov}. We confirm that the Lyapunov exponent becomes positive at a certain $a_{\rm ave}$, and thus the spatial dynamics exhibits a chaos transition. Therefore, the chaos transition and the associated breakdown of the bulk--edge correspondence are also robust against disorders. We note that the periodic solutions after the chaos transition, i.e., the windows of chaos disappear under the existence of disorder. Such disorder--induced chaos is consistent with previous research \cite{Mayer1981} on the discrete nonlinear dynamics with noises.

In some nonlinear models, eigenvalues of topological edge modes can be shifted from zero even without disorders \cite{Tuloup2020}. In contrast, the present result indicates that the nonlinear eigenvalue of edge modes in the nonlinear SSH model is zero and immune to disorders and nonlinearity. This is because of the sublattice symmetry (Eq.~(18) in Methods), which guarantees the symmetry of the band structure and the zero eigenvalue of edge modes (at least in the linear case).

\subsection{Stability of edge modes in nonlinear topological mechanics.}
While we have discussed the destabilization of edge modes by chaos transitions in the main text, some nonlinear topological insulators do not exhibit such chaos transitions at arbitrary strength of the nonlinearity. We here analyze nonlinear topological mechanics composed of elliptic gears that are studied in Ref.~\cite{Ma2023}, and show that the absence of chaos transitions is related to the linear stability in the spatial dynamics. The dynamics of that nonlinear topological mechanics is described as follows:
\begin{eqnarray}
i\partial_t \Psi_A(x) &=& f_{\epsilon}(\Psi_B(x)) - f_{-\epsilon}(\Psi_B(x-1)), \label{nonlinear_mechanics1}\\
i\partial_t \Psi_B(x) &=& f_{\epsilon}(\Psi_A(x)) - f_{-\epsilon}(\Psi_A(x+1)), \label{nonlinear_mechanics2}
\end{eqnarray}
where $\Psi_{A,B}(x)$ are the complex field variable. $f_{\epsilon}(\Psi_{A,B}(x))$ is the nonlinear function defined as
\begin{eqnarray}
 f_{\epsilon}(\Psi) &=& s_{\epsilon}({\rm Re}\,\Psi ) +  i s_{\epsilon}({\rm Im}\,\Psi ), \\
 s_{\epsilon}(x) &=& a(1-{\epsilon}^2) \int_0^x \frac{\sqrt{1+2{\epsilon}\cos x'+{\epsilon}^2}}{(1+{\epsilon}\cos x')^2} dx',
\end{eqnarray}
with $a$ being the semi major--axis of the elliptic gears and ${\epsilon}$ being their eccentricity.

\begin{figure*}[t]
  \includegraphics[width=120mm,bb=0 0 778 286,clip]{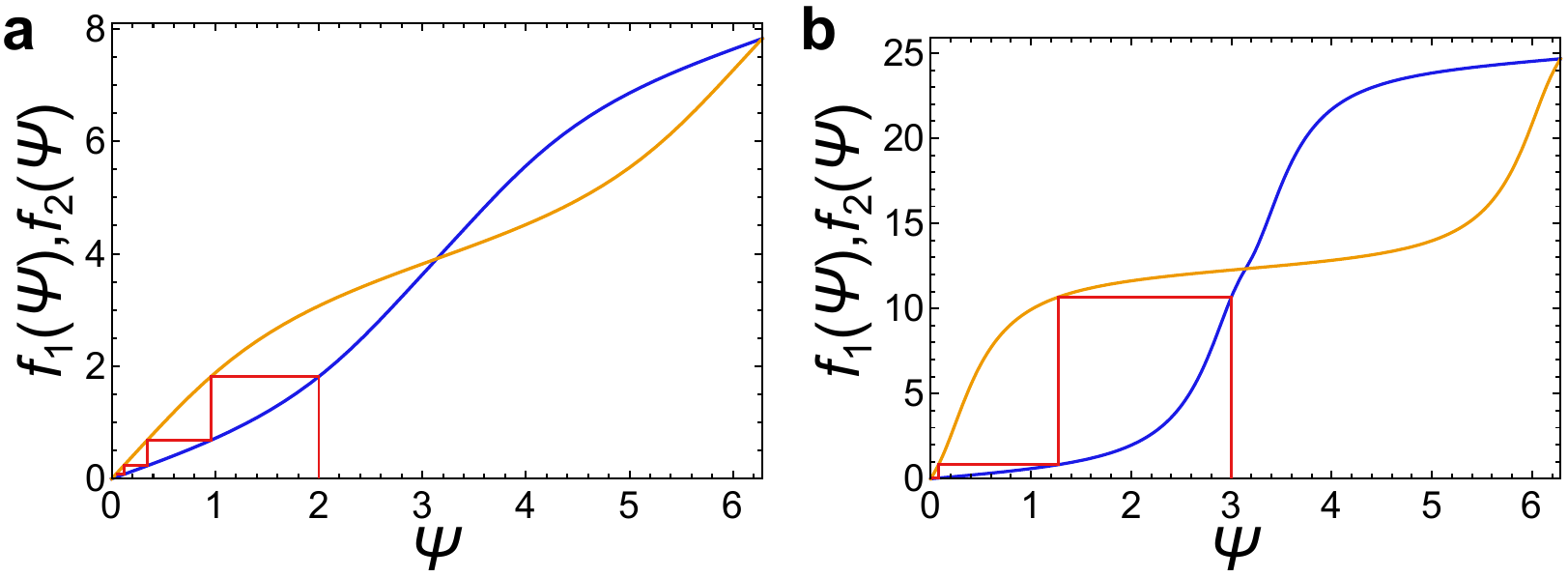}
\caption{\label{supplefig_cobweb} {\bf Cobweb plots of the spatial dynamics of a topological mechanics.} We present the cobweb plots of the spatial dynamics in Supplementary Eq.~\eqref{spatial_dynamics_mechanics}. The orange (resp. blue) curves show $y=f_{\epsilon}(x)$ (resp. $y=f_{-{\epsilon}}(x)$). The red polylines represent the spatial dynamics. If we start from $\Psi_A(1)<\pi$, $\Psi_A(x)$ must be converged to zero at any $e$, which indicates the parameter independence of the stability of the fixed point at $\Psi_A(x)=0$. We fix the parameter $a=1$ and use ${\epsilon}=0.5$ in panel {\bf a} and ${\epsilon}=0.9$ in panel {\bf b}. 
}
\end{figure*}

As in Eq.~(7) in the main text, we focus on the zero nonlinear eigenvectors and derive the nonlinear dynamics describing their spatial distribution. Assuming $\Psi_B(x)=0$, one can determine $\Psi_A(x+1)$ from $\Psi_A(x)$ so that they satisfy \begin{eqnarray}
f_{\epsilon}(\Psi_A(x)) = f_{-{\epsilon}}(\Psi_A(x+1)). \label{spatial_dynamics_mechanics}
\end{eqnarray}
Such spatial dynamics are captured by cobweb plots in Supplementary Fig.~\ref{supplefig_cobweb}. Fortunately, the value of $\Psi_A(x+1)$ is uniquely determined at any parameters.

We investigate the spatial dynamics of zero modes at different ${\epsilon}$ and find that the fixed point at $\Psi_A(x)=0$ is always stable unless the bulk topology is changed. We confirm this from the fact that $\Psi_A(x+1)$ is smaller than $\Psi_A(x)$ for $\Psi_A(x)<\pi$ and ${\epsilon}>0$. Therefore, the zero mode must be decayed to zero in the thermodynamic limit $x\rightarrow\infty$, which indicates the absence of the chaos transition. The situation is similar in other nonlinear models in, e.g., Ref.~\cite{Ma2023}, where the nonlinearity--induced topological modes are linearly stable at arbitrary parameters.

\subsection{Linear stability analysis of the fixed points in the dynamical system describing zero modes of the extended nonlinear SSH model.} While in the main text, we have numerically calculated the flow fields and demonstrated the correspondence between the nonlinear winding number and the dimension of the stable manifold in the dynamical system of zero modes, we can also calculate the dimension of the stable manifold from the linear stability analysis. Specifically, we consider the linearization around the fixed point $\Psi_{\rm fix}$ of the dynamical system in Eq.~(9) in the main text and obtain
\begin{equation}
 \left(
  \begin{array}{c}
   \delta\Psi_1(x+1)\\
   \delta\Psi_1(x)
  \end{array}
  \right) = 
  \left(
  \begin{array}{cc}
   -d/\alpha & -(a+3b|\Psi_{\rm fix}|^2)/\alpha \\
   1 & 0
  \end{array}
  \right)
  \left(
  \begin{array}{c}
   \delta\Psi_1(x)\\
   \delta\Psi_1(x-1)
  \end{array}
  \right),
  \label{extended_SSH_linearization}
\end{equation}
where $\delta \Psi = \Psi (x) - \Psi_{\rm fix}$ is the deviation from the fixed point. Then, we calculate the eigenvalues $\lambda_{\pm}$ of the matrix in this equation. The dimension of the stable manifold is equal to the number of the eigenvalues that satisfy $|\lambda_{\pm}| < 1$.

Since the dynamical system in Eq.~(9) in the main text can have three qualitatively different fixed points and limit cycles at $|\psi_{\rm fix}|^2 = 0$, $(d-a-\alpha)/b$, and $(\alpha-a)/b$, we below calculate the linear stability of these fixed points. We note that the nonzero fixed points, $|\psi_{\rm fix}|^2 = (d-a-\alpha)/b$ and $(\alpha-a)/b$, correspond to the topological edge modes that appear in the nonlinearity--induced topological phase because they converge to nonzero amplitude in the limit of $x\rightarrow\infty$. Below we focus on the parameter region $a,d,\alpha>0$, while we can conduct a similar analysis in the cases of different choices of the signs of the parameters.

\subsubsection{Stability analysis of the fixed point at $\psi_{\rm fix} = 0$.}
The linearized matrix (Supplementary Eq.~\eqref{extended_SSH_linearization}) around $\psi_{\rm fix} = 0$ exhibits the eigenvalues
\begin{equation}
\lambda_{\pm} = \frac{-d\pm\sqrt{d^2-4\alpha a}}{2\alpha}. \label{longrange_linearized_eigenvalues_weaknonlinear}
\end{equation}
To investigate the linear stability of the fixed point at $\psi_{\rm fix} = 0$, we separately consider the cases of $d^2-4\alpha a< 0$ and $d^2-4\alpha a > 0$.

First, if $d^2-4\alpha a$ is negative, the eigenvalues become a pair of complex conjugates. Then, we can calculate their absolute values as
\begin{equation}
|\lambda_{\pm}^2| = \frac{d^2+(4\alpha a-d^2)}{(2\alpha)^2} = \frac{a}{\alpha}.
\end{equation}
Thus, one can determine the linear stability of the fixed point by comparing $a$ and $\alpha$. In the case of $a<\alpha$, the fixed point is linearly stable and has a two--dimensional stable manifold. In this parameter region, the winding number becomes $\nu_{\rm NL} = 2$, which corresponds to the dimension of the stable manifold. In contrast, when $a$ is larger than $\alpha$, the fixed point is fully unstable. Then, the winding number becomes $\nu_{\rm NL} = 0$, which is also equal to the dimension of the stable manifold.

Second, if $d^2-4\alpha a$ is positive, the eigenvalues $\lambda_{\pm}$ become real. Then, we can check that $|\lambda_{-}|<1$ is equivalent to $d^2-4\alpha a < (2\alpha -d)^2$ and $2\alpha - d > 0$. We can also confirm that $|\lambda_{+}|>1$ is equivalent to $d^2-4\alpha a < (d - 2\alpha)^2$ and $d - 2\alpha > 0$. The winding number becomes $\nu_{\rm NL} = 1$ in the parameter region $d>a+\alpha$. In this case, we can show 
\begin{equation}
 (d - 2\alpha)^2 = d^2 - 4\alpha d +4\alpha^2 < d^2 - 4\alpha (a+\alpha) + 4\alpha^2 = d^2 - 4\alpha a,
\end{equation}
and thus confirm $|\lambda_{-}|>1$ and $|\lambda_{+}|<1$, which indicates that the stable manifold is one--dimension.. In contrast, if we consider the parameter region $d<a+\alpha$, we obtain $d^2-4\alpha a < (d - 2\alpha)^2$. Then, the sign of $2\alpha - d$ is determined by the sign of $\alpha - a$ as shown from the inequality 
\begin{equation}
 \alpha - a < 2\alpha - d < 2\sqrt{\alpha}(\sqrt{\alpha} - \sqrt{a}).
\end{equation}
Therefore, the dimension of the stable manifold is two in the case of $\alpha > a$ and zero in the case of $\alpha < a$. These results are consistent with the bulk--edge correspondence between the winding number and the dimension of the stable manifold.

\subsubsection{Stability analysis of the fixed point at $|\psi_{\rm fix}|^2 = (d-a-\alpha)/b$.}
If we consider the parameter region $d < a + \alpha$, $\alpha < a$, and $b<0$, we obtain the fixed point at $|\psi_{\rm fix}|^2 = (d-a-\alpha)/b$. This fixed point corresponds to the topological edge modes that emerge by the nonlinearity--induced topological phase transition at $w=(d-a-\alpha)/b$. In this case, the eigenvalues of the matrix in Supplementary Eq.~\eqref{extended_SSH_linearization} become
\begin{equation}
\lambda_{\pm} = \frac{-d\pm \sqrt{d^2+4\alpha(2a-3d+3\alpha)}}{2\alpha}.
\end{equation}
We can show that these eigenvalues are real. Since the nonlinear winding number becomes $\nu_{\rm NL} = 1$ at $w>(d-a-\alpha)/b$, we expect that the dimension of the stable manifold of the fixed point at $|\psi_{\rm fix}|^2 = (d-a-\alpha)/b$ is one. However, we can confirm the breakdown of such bulk--edge correspondence by the period--doubling bifurcation as in the original nonlinear SSH model in Eqs.~(4) and (5) in the main text.

We first check $|\lambda_-|>1$ from
\begin{equation}
(2\alpha-d)^2-[d^2+4\alpha(2a-3d+3\alpha)] = 8d\alpha - 8a\alpha - 8\alpha^2 < 0. \label{linear_stability_nonlinearSSH_NITPT1}
\end{equation}
Thus, the dimension of the stable manifold is one if $|\lambda_+| < 1 $ and otherwise zero. We first show $\lambda_+>-1$, which is equivalent to
\begin{equation}
d-2\alpha < \sqrt{d^2+4\alpha(2a-3d+3\alpha)}.
\end{equation}
We can show this inequality from 
\begin{equation}
(d-2\alpha)^2-[d^2+4\alpha(2a-3d+3\alpha)] = 8d\alpha - 8a\alpha - 8\alpha^2 < 0. \label{linear_stability_nonlinearSSH_NITPT2}
\end{equation}
Therefore, if and only if $\lambda_+$ is smaller than one, the absolute value of $\lambda_+$ is also smaller than one, i.e., $|\lambda_+|<1$. We note that $\lambda_+<1$ is equivalent to 
\begin{equation}
2\alpha + d > \sqrt{d^2+4\alpha(2a-3d+3\alpha)}.
\end{equation}
By comparing the square of the left- and right-hand sides, we obtain 
\begin{equation}
(2\alpha+d)^2-[d^2+4\alpha(2a-3d+3\alpha)] = 16d\alpha - 8a\alpha - 8\alpha^2 > 0.
\end{equation}
Therefore, $|\lambda_+|$ becomes smaller than one and thus the nonlinear winding number corresponds to the dimension of the stable manifold only in the case of $(a+\alpha)/d>2$. At the critical parameter $(a+\alpha)/d=2$, the dynamical system exhibits a period--doubling bifurcation. Such bifurcation induces the breakdown of the bulk--edge correspondence as in the original nonlinear SSH model in Eqs.~(4) and (5) in the main text.

\subsubsection{Stability analysis of the fixed point at $|\psi_{\rm fix}|^2 = (\alpha-a)/b$.}
If we consider the case of $2\alpha > d$ and $\alpha<a$, we obtain a limit cycle described as $\psi(x)=\sqrt{(\alpha-a)/b} \exp(i(\Omega x+\theta))$ whose frequency $\Omega$ satisfies $\Omega=-2\alpha\cos\theta$. This limit cycle corresponds to the topological edge mode that emerges by the nonlinearity--induced topological phase transition at $w=(\alpha-a)/b$, and the nonlinear winding number is changed into $\nu_{\rm NL} = 2$ at the transition point. The eigenvalues of the linearized matrix in Supplementary Eq.~\eqref{extended_SSH_linearization} become
\begin{equation}
\lambda_{\pm} = \frac{-d\pm\sqrt{d^2+4\alpha(2a-3\alpha)}}{2\alpha}.
\end{equation}

If we consider the case that $\lambda_{\pm}$ is complex, i.e., $d^2+4\alpha(2a-3\alpha)<0$, the absolute values of the eigenvalues satisfy
\begin{equation}
|\lambda_{\pm}^2| = \frac{d^2+[4\alpha (3\alpha-2a)-d^2]}{(2\alpha)^2} = \frac{3\alpha-2a}{\alpha} < 1.
\end{equation}
Therefore, the dimension of the stable manifold is two and corresponds to the nonlinear winding number. In contrast, if the eigenvalues are real, $|\lambda_-|<1$ is satisfied only at $a+d/2<2\alpha$, which we derive from the condition that
\begin{equation}
(2\alpha-d)^2 - [d^2+4\alpha(2a-3\alpha)] = 16\alpha^2 - 4d\alpha - 8a\alpha
\end{equation}
must be positive. We also confirm that $|\lambda_+|<1$ is equivalent to 
\begin{eqnarray}
2a-3\alpha &{\leq}& 0, \\
\sqrt{d^2+4\alpha(2a-3\alpha)} &<& 2\alpha+d.
\end{eqnarray}
Thus, we obtain $|\lambda_+|<1$ only at $a-d/2 < 2 \alpha$. If we consider the case of $a+d/2>2\alpha$, the dimension of the stable manifold is less than two, which indicates the breakdown of the bulk--edge correspondence. This breakdown is also induced by the bifurcation of the fixed point. Therefore, the breakdown of the bulk--edge correspondence by the chaos transition is independent of the hopping range in nonlinear models.

\subsection{Increased number of anti--localized modes in the long--range nonlinear SSH model.}
We here show that long--range hopping can also increase the number of anti--localized zero modes as well as topological localized modes. As discussed in the main text, we can regard the dimension of the stable manifold as the effective number of nonlinear zero eigenvectors. In the long--range nonlinear SSH model (Eq.~(9) in the main text), nonzero fixed points can have a two--dimensional stable manifold, which implies the existence of two independent anti--localized zero modes (this also indicates the existence of two localized zero modes).

\begin{figure*}[t]
  \includegraphics[width=120mm,bb=0 0 924 412,clip]{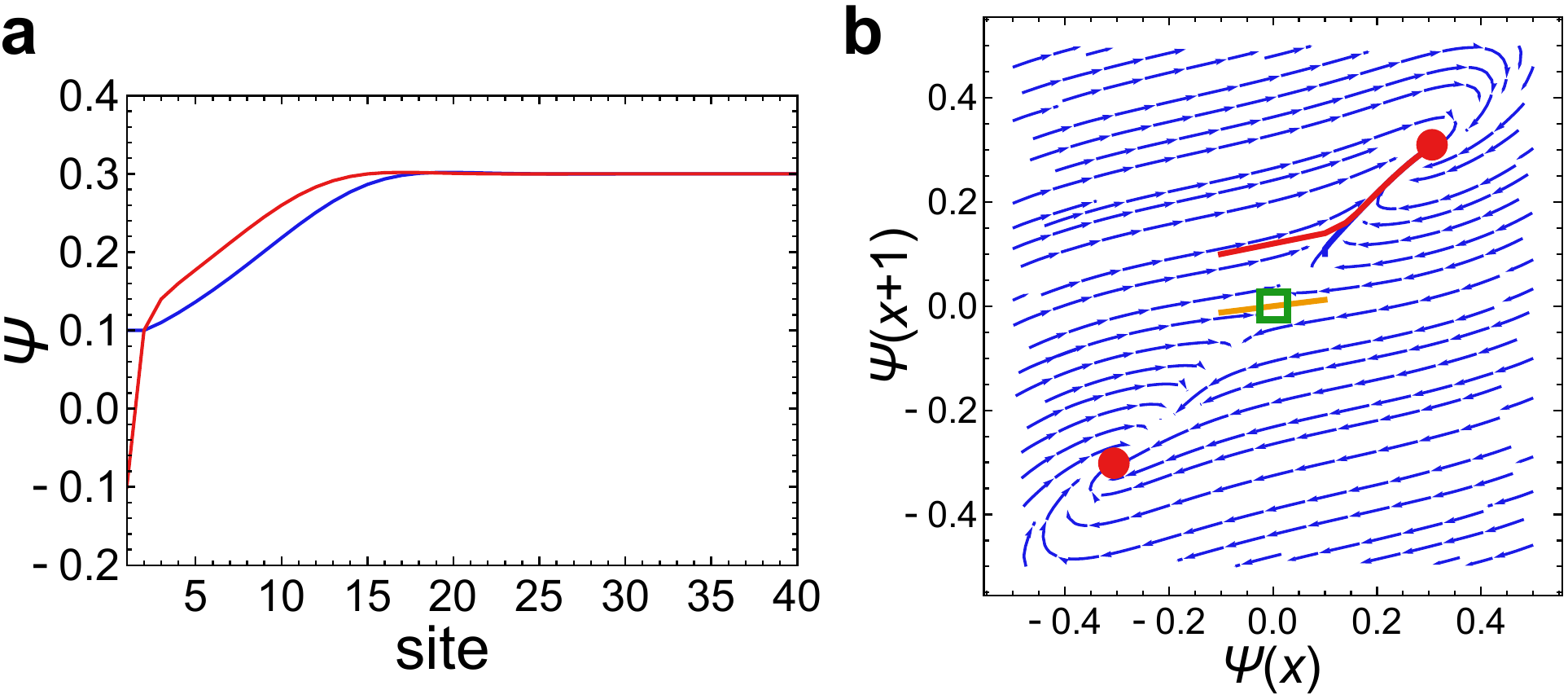}
\caption{\label{supplefig_antilocalize} {\bf Two anti--localized zero modes in the long--range nonlinear SSH model.} {\bf a} Spatial distribution of two anti--localized zero modes are presented by the red and blue polylines. By comparing the absolute value of the zero modes at the edge $x=1$ with that in the bulk $x\rightarrow \infty$, we judge their anti--localization. These anti--localized zero modes have different values at the edge, and thus we distinguish them. We use the parameters $a=0.11$, $b=c=1$, $d=-1$, and $\alpha=0.8$. {\bf b} The flow of the nonlinear dynamical system of the long--range nonlinear SSH model (Eq.~(9) in the main text) and the orbits corresponding to anti--localized zero modes. The blue curved arrows show the flow of the nonlinear dynamical system in the spatial direction. The green square and red disks are the fixed points, where the colors represent the dimensions of their stable manifolds: one (two) dimensions for the green circle (red squares). One can estimate that the red squares have two--dimensional stable manifolds from the flow around them; starting from any neighbor point, the orbit converges to the fixed point. On the other hand, the perturbation around the green fixed point diverges except for one certain direction presented by the orange line, and thus the fixed point at origin has a one--dimensional stable manifold. The blue and red curves show the orbits corresponding to the anti--localized zero modes in panel {\bf a}. Since neither of the curves is a part of the other, they correspond to independent localized modes and indicate that the values of anti--localized modes at the edge $\Psi(1)$ have two degrees of freedom. We use the same parameters as in panel {\bf a}.
}
\end{figure*}

We numerically demonstrate the existence of two anti--localized zero modes in the long--range nonlinear SSH model. We consider the parameters $a=0.11$, $b=c=1$, $d=-1$, and $\alpha=0.8$. In this case, the nonlinear winding number becomes $\nu(w)=1$ at the amplitude $w<0.09$, and $\nu(w) = 2$ at $w>0.09$. We calculate the spatial distribution of zero modes by using the nonlinear dynamical system in Eq.~(9) in the main text, starting from two different initial conditions $(\Psi_A(1),\Psi_A(2)) = (0.1,\pm 0.1)$. From both of the initial conditions, $\Psi_A(x)$ converges to $\Psi_A(x)\rightarrow 0.3$, which is larger than the initial amplitude $0.1$ (cf.~Supplementary Fig.~\ref{supplefig_antilocalize}a). Therefore, we obtain two anti--localized zero modes. This increase in the number of anti--localized zero modes corresponds to the two--dimensional stable manifold of the fixed point at $(\Psi_A(x),\Psi_A(x+1)) = (0.3,0.3)$ in Supplementary Fig.~\ref{supplefig_antilocalize}b (two dimensions of the stable manifold are analytically shown in Supplementary Note 7).

\subsection{Correspondence between the nonlinear winding number and the eigenvalues of a state-dependent transfer matrix.}
We have also considered more general nonlinear systems whose zero modes are described by Eq.~(11) in the main text. We here show that the eigenvalues of $T(g(\boldsymbol{\psi}_1(x)))$ in Eqs.~(11) and (12) in the main text are related to the nonlinear winding number, which is the origin of the bulk--edge correspondence. Specifically, if we consider a fixed value of $g(\boldsymbol{\psi}(k))=w$ independently of the wavenumber $k$, the nonlinear winding number 
\begin{equation}
 \nu(w) = \frac{1}{2\pi i} \int_0^{2\pi} \partial_k \log \det \left( A(w)+e^{ik}D(w) \right) dk. \label{winding_num2}
\end{equation}
is equal to the number of the eigenvalues of $T(w)=D^{-1}(w) A(w)$ whose absolute values are less than one.

To prove the correspondence between the nonlinear winding number and the eigenvalues of the transfer matrix $T(w)$, we explicitly write down the eigenequation of $T(w)$:
\begin{equation}
\det(T(w)- \lambda I) = \det(D^{-1}(w) A(w) - \lambda I) = 0 \label{eigeneq_transfer_mat}
\end{equation}
with $\lambda$ being the eigenvalue and $I$ being an identity matrix. Since we assume that $D$ is a regular matrix, Supplementary Equation \eqref{eigeneq_transfer_mat} is equivalent to 
\begin{equation}
\det(A(w) - \lambda D(w)) = 0. \label{eigeneq_transfer_mat2}
\end{equation}
Meanwhile, the nonlinear winding number in Supplementary Eq.~\eqref{winding_num2} is the winding number of $\det \left( A(w)+zD(w) \right)$, when we move $z$ along a unit circle in a complex plane. According to the argument principle in complex analysis, such a winding number corresponds to the number of solutions of the algebraic equation $\det \left( A(w)+zD(w) \right)=0$ that satisfies $|z|<1$. By relating $\lambda$ in Supplementary Eq.~\eqref{eigeneq_transfer_mat2} and $z$, these arguments indicate that the nonlinear winding number is equal to the number of eigenvalues of $T(w)$ whose absolute values are less than one.

\subsection{Possible $\mathbb{Z}\times\mathbb{Z}$ classification in the nonlinear SSH model.} If we consider the case of $b\neq c$ in the nonlinear SSH model (Eqs.~(4) and (5) in the main text), edge modes localized at the left and right boundaries can appear at different critical amplitudes. To show that, we consider both left and right semi--infinite systems, $x\leq -1$ and $x \geq 1$, respectively. While the spatial distribution of zero modes in the right semi--infinite system is described by 
\begin{equation}
\Psi_A(x+1) = -\frac{a+b|\Psi_A(x)|^2}{d} \Psi_A(x), \label{edge_dynamics_right}
\end{equation}
(the same as Eq.~(7) in the main text), that in the left semi--infinite system is described by 
\begin{equation}
\Psi_B(x-1) = -\frac{a+c|\Psi_B(x)|^2}{d} \Psi_B(x), \label{edge_dynamics_left}
\end{equation}
and $\Psi_A(x)=0$. The difference between $b$ and $c$ leads to the different transition points in Supplementary Eqs.~\eqref{edge_dynamics_right} and \eqref{edge_dynamics_left}; the edge mode in the right semi--infinite system appears at $w=(d-a)/b$, while that in the left semi--infinite system appears at $w=(d-a)/c$. Therefore, the left edge mode and right edge mode have different critical amplitudes.

The difference in the critical amplitudes indicates the possible $\mathbb{Z}\times\mathbb{Z}$ classification in the nonlinear Su-Scrieffer-Heeger (SSH) model because the numbers of left and right edge modes can correspond to different topological invariants. In fact, one can define two nonlinear winding numbers in the nonlinear SSH model,
\begin{eqnarray}
\nu_{l} &=& \frac{1}{2\pi i} \int_0^{2\pi} dk \partial_k \log (a+bw+de^{ik}), \\
\nu_{r} &=& \frac{1}{2\pi i} \int_0^{2\pi} dk \partial_k \log (a+cw+de^{ik}),
\end{eqnarray}
whose transition points are consistent with the amplitude where the left and right edge modes appear. Such a $\mathbb{Z}\times\mathbb{Z}$ classification may be analogous to that in non-Hermitian systems, where the point-gap topology of one--dimensional systems is also classified by $\mathbb{Z}\times\mathbb{Z}$ topological invariants under the existence of the sublattice symmetry \cite{Kawabata2019}. However, the nonlinear SSH model considered here is a conservative system and preserves the energy in its time evolution. Therefore, nonlinearity in itself can induce non-Hermitian-like effects in the classification of topology.

We can also consider the change of the definition of $w$ to elucidate the difference of the parameter $w$ where the left and right edge modes appear. We discuss the possibility of different definitions of $w$ and the extension of the bulk--edge correspondence for such $w$ in the following section.

\subsection{Change of definition of the parameter $w$ according to the nonlinearity in more general systems.}
In the main text, we have considered the nonlinear SSH model whose nonlinear term is proportional to $|\Psi_A(x)|^2+|\Psi_B(x)|^2$. In general, if the nonlinear dynamics is described as
\begin{equation}
i\partial_t \Psi_j(x) = \sum_{l,x'} H_{jl}\left(x,x';\sum_m |\Psi_m(x)|^2\right) \Psi_l(x'),
\end{equation}
where $H$ is a matrix parametrized by $\sum_m |\Psi_m(x)|^2$, the nonlinear eigenvalue problem is equivalent to a linear eigenvalue problem under the fixed $\sum_m |\Psi_m(x)|^2=w$. Then, one can exactly calculate the nonlinear winding number from the corresponding linear winding number.

Since the amplitude $\sum_m |\Psi_m(x)|^2=w$ is unchanged under the time evolution in a conserved system, it is quite natural to focus on the special solutions with the fixed amplitude. However, if the strengths of nonlinear terms are not determined by the amplitude, there can be other choices of $w$ to define the nonlinear winding number. Specifically, if the nonlinear dynamics is described as
\begin{equation}
i\partial_t \Psi_j(x) = \sum_{l,x'} H_{jl}\left(x,x';g(\Psi(x))\right) \Psi_l(x'), \label{general_nonlinear_dynamics}
\end{equation}
where $g(\Psi(x))$ is a nonlinear function of $\Psi_j(x)$ ($j=1,\cdots,M$ with $M$ being the internal degree of freedom), and $H$ is a matrix parametrized by $g(\Psi(x))$, it may be better to fix $g(\Psi(x))$ as $g(\Psi(x))=w$ instead of the amplitude.

We first show that fixing $g(\Psi(x))=w$ enables us to calculate the nonlinear eigenequation (Supplementary Eq.~\eqref{general_nonlinear_dynamics}). The wavenumber--space description of the nonlinear eigenequation corresponding to Supplementary Eq.~\eqref{general_nonlinear_dynamics} becomes
\begin{equation}
E(k) \psi_j(k) = \sum_{l} H_{jl}\left(k;g(\psi(k))\right) \psi_l(k). \label{general_nonlinear_dynamics_wavenumber}
\end{equation}
If we focus on special solutions with $g(\psi(k))=w$ being fixed independently of $k$, the right-hand side reads $\sum_{l} H_{jl}\left(k;w \right) \psi_l(k)$. Therefore, the nonlinear eigenvalue problem becomes equivalent to the linear eigenvalue problem parametrized by $w$ and $k$, and thus one can calculate the eigenvalues and eigenvectors by solving linear equations. We note that the multiple of the obtained eigenvector $\psi(k)$ is also an eigenvector because of the linearity. Thus, the final step of the calculation of the nonlinear eigenvector is to find the constant $c(k)$ that satisfies 
\begin{equation}
g(c(k)\psi(k))=w. \label{equation_w}
\end{equation}
Such $c(k)$ can be absent for certain $\psi(k)$ and $w$, while one can guarantee the existence of $c(k)$ satisfying Supplementary Eq.~\eqref{equation_w} if the nonlinear function has desirable properties. Specifically, if $g$ is a continuous function, one can always find $c(k)$ for $w$ in the range of $\max_c \min_{\psi} g(c\psi) \leq w \leq \min_c \max_{\psi} g(c\psi)$.

We can also discuss the bulk--edge correspondence of the winding number under the fixed $f(\Psi(x))=w$. In particular, if we focus on the generalized nonlinear SSH model,
\begin{eqnarray}
i\partial_t \Psi_A(x) &=& (a+g(\Psi_A(x),\Psi_B(x))) \Psi_B(x) + d \Psi_B(x-1), \label{generalized_nonlinearSSH1}\\
i\partial_t \Psi_B(x) &=& (a+g(\Psi_A(x),\Psi_B(x))) \Psi_A(x) + d \Psi_A(x+1), \label{generalized_nonlinearSSH2}
\end{eqnarray}
where $g(\Psi_A(x),\Psi_B(x))$ is a real nonlinear function of $\Psi_A(x)$ and $\Psi_B(x)$, we can confirm the bulk--edge correspondence in a similar sense to that in the original nonlinear SSH model (Eqs.~(4) and (5) in the main text). First, we can calculate the nonlinear winding number as $\nu(w)=\int_0^{2\pi} dk \partial_k \log (a+w+e^{ik}) / (2\pi i)$, which is $\nu=1$ (resp. $\nu=0$) in the case of $a+w<d$ (resp. $a+w>d$). Then, we derive the nonlinear dynamical system describing the spatial distribution of zero modes in the semi--infinite system,
\begin{equation}
\Psi_A(x+1) = -\frac{a+g(\Psi_A(x),0)}{d} \Psi_A(x). \label{edge_dynamics_genralized}
\end{equation}
When $a+g(\Psi_A(x),0)$ is smaller (larger) than $d$, $|\Psi_A(x+1)|$ becomes smaller (larger) than $|\Psi_A(x)|$. This implies that there is a fixed point at $\Psi'_A(x)$ satisfying $a+g(\Psi'_A(x),0)=d$, and it can be stable if $\nu(w)$ is one for at $w=g(\Psi'_A(x)+\delta,0)$ and zero at $w=g(\Psi'_A(x)-\delta,0)$ with $\delta$ being a sufficiently small constant. If $\Psi'_A(x)$ is a stable fixed point, there are nonlinear edge modes with the initial amplitude $\Psi_A(1)$ satisfying $\nu(g(\Psi_A(1),0))=1$, which indicates the bulk--edge correspondence of the nonlinear winding number. However, Supplementary Equation \eqref{edge_dynamics_genralized} can exhibit bifurcations, and $\Psi'_A(x)$ can be an unstable fixed point after the bifurcations. Therefore, the bulk--edge correspondence is only valid for sufficiently weak nonlinearity as in the original nonlinear SSH model.



\end{document}